\documentclass[a4paper,10pt,notitlepage]{article}
\usepackage{amsfonts}
\usepackage{amsmath,amsthm}

\numberwithin{equation}{section}

\newtheorem{lemma}{Lemma}[section]

\newcommand{\beqa}{\begin{eqnarray}}
\newcommand{\eeqa}{\end{eqnarray}}
\DeclareMathOperator{\str}{\text{str}}

\newcommand{\calH}{\mathcal{H}}
\newcommand\Ss{\ensuremath{S^{3|2}}}
\newcommand\Ssn{\ensuremath{S^{2S+1|2S}}}

\newcommand\osp{\ensuremath{\text{osp}(4|2)}}
\newcommand\aosp{\ensuremath{\widehat{\text{osp}}(4|2)}}
\newcommand\OSPn{\ensuremath{\text{OSP}(2S+2|2S)}}
\newcommand\ospn{\ensuremath{\text{osp}(2S+2|2S)}}
\newcommand\aospn{\ensuremath{\widehat{\text{osp}}(2S+2|2S)}}
\newcommand\tr{{\text{tr}}}
\newcommand{\Integer}{{\mathbb{Z}}}
\newcommand\wosp{\ensuremath{\widehat{\text{osp}}(4|2)_{1}}}
\newcommand\wospn{\ensuremath{\widehat{\text{osp}}(2S+2|2S)_{1}}}
\newcommand\fu{\text{f}}
\newcommand\Sa{{\cal S}}
\newcommand\cJ{{\cal J}}
\newcommand\Sp{S}

\title{\textbf{Principal Chiral Model on Superspheres}}

\author{\\[5mm]Vladimir Mitev$^1$, Thomas Quella$^2$ and Volker
  Schomerus$^1$\\[5mm]
$^1$ DESY Hamburg, Theory Group, \\
Notkestrasse 85, D--22607 Hamburg, Germany
\\[5mm]
$^2$ Institute for Theoretical Physics, University of Amsterdam,\\
Valckenierstraat 65, 1018 XE Amsterdam,
The Netherlands}
\date{}

\begin{document}
\begin{titlepage}
    \maketitle
    \begin{abstract}
       We investigate the spectrum of the principal chiral model (PCM) on
        odd-dimensional superspheres as a function of the curvature radius
        $R$. For volume-filling branes on \Ss, we compute the exact
        boundary spectrum as a function of $R$. The extension to higher
        dimensional superspheres is discussed, but not carried out in
        detail. Our results provide very convincing evidence in favor of
        the strong-weak coupling duality between supersphere PCMs and
        \OSPn\ Gross-Neveu models that was recently conjectured by
        Candu and Saleur.
    \end{abstract}
\vspace*{-13cm} {\tt {DESY 08-123}} \hfill {\tt {0809.1046\\}}
{\tt {ITFA-2008-30}} \hfill {\tt {NI09007-SIS\\}}
\bigskip\vfill
\noindent \phantom{wwwx}{\small e-mail: }{\small\tt
Vladimir.Mitev@desy.de,T.Quella@uva.nl, Volker.Schomerus@desy.de}
\end{titlepage}

\tableofcontents

\section{Introduction}
\addtolength{\baselineskip}{1mm}

Principal chiral models (PCMs) on symmetric spaces have been
studied extensively because of their numerous applications in many
different branches of physics. While PCMs on symmetric spaces are
well-known to possess an infinite number of classically conserved
quantities (see \cite{Pohlmeyer:1975nb,Luscher:1977rq,Luscher:1977uq,
Brezin:1979am,Eichenherr:1979ci,Eichenherr:1981sk} for early work and
e.g.\ \cite{Schwarz:1995td,Evans:2000qx} for more recent developments 
and references), quantum effects spoil integrability in many cases
\cite{Goldschmidt:1980wq,Abdalla:1982yd}. And even in those examples 
for which this does not happen, finding explicit formulas for
partition functions and correlators is a difficult problem that
has only been solved for a small set of models. More recently,
PCMs on (generalized) symmetric {\em super}spaces have received
considerable attention. This is explained in part through the 
role they play for the description of strings for 
Anti-de\,Sitter (AdS) backgrounds in various dimensions, 
including $AdS_5\times S^5$ and $AdS_4\times\mathbb{C}\mathbb{P}^3$
\cite{Metsaev:1998it,Berkovits:2000fe,Arutyunov:2008if,Fre:2008qc}.
PCMs on symmetric superspaces possess a number of remarkable 
properties. In particular, there exist several families 
of quantum {\em conformal} models \cite{Berkovits:1999im,Bershadsky:1999hk,Zirnbauer:1999ua,Read:2001pz,Babichenko:2006uc}.
Yet, finding explicit solutions is still rather difficult and will 
certainly require developing new techniques, see e.g.\ \cite{Mann:2004jr,
Mann:2005ab}. Some remarkable recent advances, most importantly the
results of \cite{Quella:2007sg} and \cite{Candu:2008vw,Candu:2008yw},
seem to bring at least some partial solutions within reach. One of
our aims here is to initiate and explore new solution strategies
that incorporate target space supersymmetry as an essential feature.

In this work we focus on a particular family of symmetric target
superspaces, namely on the odd dimensional superspheres \Ssn\ with
$2S$ fermionic coordinates. The supersphere \Ssn\ admits at least
three different descriptions that will be somewhat useful for us
below. We can think of \Ssn\ as a supermanifold in
$\mathbb{R}^{2S+2|2S}$ defined by the equation
\begin{equation}\label{constraint}
\sum_{i=1}^{2S+2}  x^2_i + 2 R^2 \sum_{a=1}^S\eta_{2a-1} \eta_{2a}
\ = \ R^2\ \ .
\end{equation}
Here, $x_i, i = 1, \dots,2S+2,$ and $\eta_j, j=1,\dots,2S,$ are
the bosonic and fermionic coordinates of $\mathbb{R}^{2S+2|2S}$,
respectively. The real parameter $R$ has been introduced to denote
the radius of the supersphere. Note that in our conventions, the
bosonic coordinates scale with the length while the fermionic
coordinates are chosen to be dimensionless. From our description
of the supersphere through equation \eqref{constraint} it is
evident that \Ssn\ comes equipped with an \ospn\ action. In fact,
the Lie superalgebra \ospn\ acts on the embedding space
$\mathbb{R}^{2S+2|2S}$ through its fundamental representation. By
the very definition of \OSPn\ this action respects the constraint
\eqref{constraint}. Hence, we arrive at a second description of
\Ssn\ as a symmetric space
\begin{equation}\label{coset}
  \Ssn \ = \ \OSPn/\text{OSP}(2S+1|2S)\ \ .
\end{equation}
Note that the stabilizer of any point on the supersphere is
isomorphic to the subsupergroup $\text{OSP}(2S+1|2S)
\subset\OSPn$. Finally, we can also solve the
constraint~\eqref{constraint} explicitly by parametrizing the
supersphere \Ssn\ through $2S+1$ angular coordinates $\varphi_j$
and $2S$ fermionic variables $\eta_j$. In the case of the 3-sphere
\Ss, for example, the line element takes the following form
\begin{equation}\label{metric}
ds^2 \ = \ 2 R^2 (1-\eta_1\eta_2) d\eta_1d\eta_2 + R^2 (1-2
\eta_1\eta_2)d\Omega_3
\end{equation}
where
$$
d\Omega_3 \ = \ d\varphi^2_1 + \cos^2 \varphi_1\ d\varphi_2^2 +
\sin^2 \varphi_1 \ d\varphi_3^2
$$
is the usual line element of the 3-dimensional unit sphere. All
three descriptions of the supersphere \Ssn\ will be used
frequently throughout the rest of this work. \medskip

Next we turn to the principal chiral model on the supersphere.
Once more, there are different ways to introduce this theory. The
most basic one is to think of it as a linear sigma model for the
fields $x_i$ and $\eta_j$ with a non-linear constraint
\eqref{constraint} on the field configurations. Another
possibility is to consider it as a non-linear sigma model. In the
case of the 3-dimensional supersphere the latter takes the form
\begin{equation} \label{PCM}
\begin{split}
\Sa^{\text{PCM}} \ = \ & \frac{R^2}{2\pi} \int d^2z \left(
 2 (1-\eta_1\eta_2) \left(\partial\eta_1 \bar \partial \eta_2
 - \partial \eta_2 \bar\partial \eta_1\right)\phantom{\frac12} \right. \\[2mm]
 & \hspace*{1cm}\left. \phantom{\frac12}
+ (1-2\eta_1\eta_2) \left(\partial \varphi_1 \bar \partial
\varphi_1 + \cos^2 \varphi_1\, \partial \varphi_2 \bar
\partial\varphi_2 + \sin^2
\varphi_1\, \partial \varphi_3 \bar \partial\varphi_3\right)\right)\\
\end{split}
\end{equation}
for the fields $\eta_j, \varphi_i$. The coupling constant in front
of the action is determined by the radius $R$ of \Ss. For the PCM
on the purely bosonic 3-sphere the coupling $R$ runs and in order
for the flow to end in a non-trivial fixed-point one must add a WZ
term \cite{Witten:1983ar}. But the presence of the two fermionic
directions changes the situation drastically. As shown in
\cite{Read:2001pz}, the $\beta$-function of the PCM on \Ssn\ is
the same as for a bosonic PCM on a sphere $\Sp^d$ whose dimension
$d=2S+1-2S=1$ is given by the difference between the number of
bosonic and fermionic coordinates. Consequently, the
$\beta$-function vanishes for the PCM on \Ssn, i.e.\ the model
\eqref{PCM} defines a family of conformal field theories at
central charge $c=1$ with continuously varying exponents. \medskip

Of course, unlike the PCM on $\Sp^1 =$ U(1), the theory defined by
the action \eqref{PCM} is not free. For large radius $R$, the
model is weakly coupled and its properties may by studied
perturbatively. But as we pass to a more strongly curved
background, computing quantities as a function of the radius $R$
may seem like a very daunting task. This is even more so because
there is very little symmetry to work with. As a conformal field
theory, the PCM on the supersphere possesses the usual chiral
Virasoro symmetries. But for a model with multiple bosonic
coordinates the two sets of chiral Virasoro generators are not
sufficient to make the theory rational. In addition, there is a
single set of global \osp\ generators. Their Noether currents,
however, fail to be chiral, at least for generic points in the
moduli space. Without the protection of current algebra
symmetries, the usual algebraic tools of conformal field theory
cannot be applied to supersphere PCMs and so we have to proceed
along a rather different route.
\medskip

Many years of experience with sigma models show that they often
possess interesting dual descriptions. The simplest such duality
is that between the free compactified boson and the massless
Thirring model. Let us recall that the latter involves two real
fermions $\psi_1$ and $\psi_2$ and the following action
$$ \Sa^{\text{Th}}_{m=0} \ = \frac{1}{2\pi} \int d^2z \sum_{i=1}^2
\biggl[
 \psi_i \bar\partial \psi_i +
  \bar{\psi}_i \partial \bar{\psi}_i
  + g^2 \bigl(\psi_1 \bar{\psi}_2 -  \psi_2 \bar{\psi}_1\bigr)^2\biggr]\
$$
where the compactification radius $R$ is related to the coupling
$g$ through $R^2 = 1+g^2 $.
Similarly, one may hope to uncover a dual description of the PCM
on the supersphere \Ssn\ that becomes weakly coupled for some
finite value of the radius $R$, deep in the strongly curved
regime. Such a dual description was indeed proposed recently.
According to an intriguing conjecture by Candu and Saleur
\cite{Candu:2008yw}, there indeed exists one special radius $R =
R_0$ at which the PCM on \Ssn\ can be described as a
non-interacting Gross-Neveu model involving $2S+2$ real fermions
$\psi_i$ along with $S$ bosonic $\beta\gamma$ systems $\gamma_a$
and $\beta_a$,
\begin{equation}\label{FFact}
\Sa^{\text{GN}}_{g=0} \ = \ \frac{1}{2\pi} \int d^2z \biggl[
 {\sum}_i \bigl(\psi_i \bar\partial \psi_i +
  \bar{\psi}_i \partial \bar{\psi}_i\bigr)
  + {\sum}_a \bigl(\beta_a \bar \partial \gamma_a + \bar{\beta}_a \partial
  \bar{\gamma}_a\bigr) \biggr]\ .
\end{equation}
All the fields appearing in this theory possess conformal weight
$h_i = h_a = 1/2$ so that the central charge is
$c=S+1-S=1$. At this point in the moduli space, the theory
possesses two commuting sets of chiral \osp\ currents $J^\mu =
J^\mu(z)$ and $\bar J^\mu = \bar J^\mu(\bar z)$. Explicit formulas
will be spelled out in section 3 below. The affine symmetry is
broken down to a global \osp\ symmetry by the following \osp\
invariant marginal deformation
\begin{equation} \label{int}
\Sa^{\text{int}} \ = \ \frac{g^2}{2\pi} \int d^2z  J_\mu(z) \Omega
(\bar J^\mu(\bar z)) \ = \ \frac{g^2}{2\pi} \int d^2z \left[{\sum}_i
\varpi_i \psi_i \bar{\psi}_i + {\sum}_a (\gamma_a
\bar{\beta}_a - \beta_a \bar{\gamma}_a)\right]^2 \ .
\end{equation}
Here, $\Omega$ is a particular automorphism of the osp(2S+2$|$2S) current
algebra which leaves a subalgebra osp(2S+1$|$2S) invariant. It
will be spelled out explicitly below. The numbers $\varpi_i$ are
given by $\varpi_1 = -1$ and $\varpi_i = 1$ for $i \neq 1$. The
theory $\Sa^{\text{GN}} = \Sa^{\text{GN}}_{g=0}+ \Sa^{\text int}$ is
claimed to be equivalent to the supersphere PCM with the two
coupling constants $R$ and $g$ related by $R^2 =
1+g^2$.\footnote{Let us note that the signs $\varpi_i$ in the
iteraction term are directly linked to the automorphism $\Omega$.
These signs were missing in the original formulation of the
conjecture by Candu and Saleur \cite{Candu:2008yw}. They are
irrelevant for $S=0$ but play a certain role when $S \geq 1$.} The
equivalence is a strong-weak coupling duality since $\Sa^{\text{GN}}$
becomes weakly coupled for $R \sim R_0 = 1 $. Note that this
duality is a direct generalization of the relation between the
compactified free field and the massless Thirring model. There
appears one real fermion for each bosonic coordinate of the
embedding space $\mathbb{R}^{2S+2|2S}$. Each pair of additional
fermionic directions gives rise to a $\beta\gamma$ system. Note,
however, that the duality between supersphere PCMs and Gross-Neveu
models is one between interacting conformal field theories. In that
sense, it is much less trivial then its purely bosonic counterpart.
\medskip

The main aim of this note is to provide very compelling evidence
for the duality between the theory (\ref{FFact},\ref{int}) and the
supersphere PCMs, extending previous numerical and algebraic 
arguments given in \cite{Candu:2008vw,Candu:2008yw}. To this end 
we shall employ some recent results of \cite{Quella:2007sg} that 
are designed to compute exact spectra in models with a special 
class of target space supersymmetries, including the two series 
psl(N$|$N) and
\ospn. The Lie superalgebra \ospn\ possesses a vanishing quadratic
Casimir $C_{\text{ad}} \sim f_{\mu\nu\rho}f^{\mu\nu\rho}$ in the
adjoint representation. Since $C_{\text{ad}}$ may be considered as
a rough measure for the `amount of non-abelianess' of a Lie
superalgebra, one may suspect that field theories with \ospn\
symmetry are somewhat intermediate between free field theories and
the most general interacting models. Indeed, as was shown in
\cite{Bershadsky:1999hk,Quella:2007sg}, the perturbation series
for conformal weights has features that are very reminiscent of
those in abelian models (torus compactifications). In this note we
shall construct the exact partition function of the theory
(\ref{FFact},\ref{int}) with a particular choice of boundary
conditions, but for all values(!) of the coupling $g$. We shall
prove that it interpolates correctly between $g = 0$ and the
spectrum of the supersphere PCM  at $R=\infty$.
\medskip

The main results of \cite{Quella:2007sg} are rather easy to state.
Before we do so, let us briefly review the behavior of conformal
weights for a compactified free bosonic field $\varphi \sim
\varphi + 2\pi R$. Suppose we are given a field $\Psi$ of
conformal weight $h_0(\Psi)$ at some radius $R_0$. In order to
find the conformal weight of the same field $\Psi$ at a different
radius $R$, it suffices to know its U(1) charge $g(\Psi)$
(momentum/winding). The conformal weight is then given by
\begin{equation} \label{dimshiftC} h(\Psi) \ =\ h_0(\Psi) + f(R)\,
g^2(\Psi)\end{equation} where $f(R)$ is some {\em universal}
function of the radius that is the same for all fields $\Psi$.
$f(R)$ may depend, however, on whether $\Psi$ is a bulk or
boundary field and on the precise boundary condition that is
imposed. For bulk fields, there exist independent left and right
U(1) charges and the behavior of the weights is a bit more
complicated. We shall briefly comment on this issue in the
conclusions. Returning to our supersphere conformal field
theories, we pick any field $\Psi$ of weight $h_0(\Psi)$ in the
free field theory \eqref{FFact}. Let us suppose that $\Psi$ is
part of some \ospn\ multiplet $\Lambda$. According to the
arguments explained in \cite{Quella:2007sg} (see also
\cite{Candu:2008vw} for numerical checks), its dimension at radius
$R$ is then given by
\begin{equation} \label{dimshiftSs}  h(\Psi) \ = \ h_0(\Psi) +
f(R)\  C_2(\Lambda) \ \ .
\end{equation}
Here, $C_2(\Lambda)$ is the value of the quadratic Casimir element
in the representation $\Lambda$ of the Lie superalgebra \ospn.
Once again, the function $f(R)$ is universal, i.e.\ it does not
depend on the field $\Psi$. Hence, the shift of the conformal
weight is entirely determined by the way $\Psi$ transforms under
the action of the Lie superalgebra \ospn. Equation
\eqref{dimshiftSs} is the direct generalization of
eq.~\eqref{dimshiftC} with the square of the U(1) charge replaced
by the quadratic Casimir. The behavior \eqref{dimshiftSs} has been 
also been predicted through the study of lattice algebras in 
\cite{Candu:2008yw}. It was furthermore checked using  
perturbative calculations at $R=\infty$ and with numerical 
simulations.  We shall refer to the behavior \eqref{dimshiftSs} 
as a quasi-abelian deformation of conformal
weights. It is typical for models with \ospn\ or psl(N$|$N)
symmetry, though often restricted to particular (boundary) fields
of the theories (see \cite{Quella:2007sg} and final section for
more details). Let us mention that fields transforming in
representations with vanishing Casimir $C_2(\Lambda)$ are
protected, i.e.\ their conformal weights are independent of $R$.
Multiplets of this type always satisfy some shortening conditions.
Our formula \eqref{dimshiftSs}, however, applies to {\em all}
fields in the theory, irrespectively of whether they are long or
short. It allows to compute their conformal weight for all values
of the radius $R$.
\medskip

Let us study a few concrete examples of the quasi-abelian
deformation of conformal weights. In the large volume limit, the
PCM possesses an infinite number of fields with conformal weight
$h = 0$. These simply correspond to functions on the supersphere.
The simplest function is the constant. Since it transforms in the
trivial representation of \ospn, its conformal weight remains
undeformed at $h=0$. It corresponds to the unique vacuum state of
the free Gross-Neveu model \eqref{FFact}. Next, the PCM contains
the fundamental multiplet $x_i,\eta_j$. The quadratic Casimir of
this multiplet $\Lambda = \Lambda_f$ is $C_2(\Lambda_f) = 1$,
i.e.\ its value is independent of $S$. As we move from the free
sigma model at $R=\infty$ towards the free Gross-Neveu model
\eqref{FFact}, the fields $x_i,\eta_j$ acquire a non-vanishing
anomalous dimension which becomes $h = h_0 + f(R_0) C_2(\Lambda_f)
= 1/2$ when we reach the radius $R = 1$ corresponding to $g=0$.
Hence, the fundamental multiplet of the PCM turns into the
multiplet $\psi_i,\gamma_a, \beta_a$. Higher functions possess
larger Casimir and hence they are mapped to states of weight $h
> 1/2$ at $g=0$. Beyond the space of ground states in the PCM,
there are fields involving any number of world-sheet derivatives.
These have positive integer weight at $R = \infty$. As we shall
see below, such states can transform in \ospn\ representations
$\Lambda$ with both positive and negative values $C_2(\Lambda)$ of
the quadratic Casimir. Consequently, some of these multiplets are
moved up while others are moved down to lower weights. Our claim
is that weights are rearranged in precisely the right way to
reproduce the spectrum of the $g=0$ Gross-Neveu model.
\medskip

The plan of this work is as follows. In the next section we shall
study the PCM \eqref{PCM} for the 3-dimensional supersphere \Ss\
and determine its exact spectrum at $R=\infty$. For simplicity, we
shall also restrict to the partition function on a strip with
Neumann boundary conditions imposed along both boundaries. After a
detailed discussion of the low lying states, we present a closed
formula for the full partition function \eqref{The spectrum}. The
latter is then decomposed explicitly into the contributions coming
from states which transform in the same representation $\Lambda$
under the global \osp. Section 3 is devoted to the theory
\eqref{FFact} and its deformation by the term \eqref{int}. In
particular, we study the bulk and boundary spectrum of the free
field theory. One of the resulting boundary partition functions is
then expanded explicitly in terms of \osp\ characters. This allows
us to compare with the spectrum of the PCM at radii $R < \infty$,
using some of the tools developed in \cite{Quella:2007sg}. We
shall find that the results agree exactly with the partition
function found in section 2! In the fourth section, we comment on
the generalization to higher dimensional superspheres. Finally,
the conclusions contain a few general thoughts on possible
implications for string theory in Anti-deSitter spaces. We shall
also briefly discuss the computation of bulk spectra for odd
dimensional superspheres.

\section{Spectrum of the supersphere PCM at large volume}

In this section we shall focus on the PCM for the supersphere \Ss\
with large radius $R$. At the point $R = \infty$ we can compute
partition functions for periodic boundary conditions and on a
strip. The two main ingredients are the exact minisuperspace
spectrum on \Ss\ (see subsection 2.1) and a good control of the
combinatorics that determine the field theoretic spectrum at
$R=\infty$. The latter will be explained in subsection 2.2. The
spectrum is finally decomposed into finite dimensional
representations of the global symmetry algebra \osp\ in the third
subsection.

\subsection{Particle on the supersphere \Ss}

The Laplacian on the supersphere \Ss\ was analyzed in full detail
by Candu and Saleur \cite{Candu:2008yw}. We shall state their
results first and then provide a new derivation that is
particularly well suited for the discussion in the following
subsections.
\smallskip%

As a warm-up, let us briefly recall the spectrum of the Laplacian
on a 3-sphere $\Sp^3$. The space of functions on $\Sp^3$ carries
an action of so(4)$ \cong $sl(2)$\oplus$sl(2). Therefore,
eigenfunctions of the Laplacian on $\Sp^3$ are organized in finite
dimensional multiplets of sl(2)$\oplus$sl(2). According to the
Peter-Weyl theory for $\text{SU}(2) \cong \Sp^3$, there is one such
multiplet $\varphi_m$ for each integer $m=0,1,2,\dots$. It has
dimension $d_m = (m+1)^2$ and transforms in the representation
$(\frac{m}{2},\frac{m}{2})$. The eigenvalue of the Laplacian on
the multiplet $\varphi_m$ is given by $\Delta_m = m(m+2)$. For the
supersphere \Ss\ we expect very similar results except that the
multiplicities should roughly exceed those of the bosonic model by
a factor of $4$.
\medskip

Before we extend these thoughts to the supersphere, however, let
us mention a few facts on the Lie superalgebra \osp. Its bosonic
subalgebra is 9-dimensional and it consists of three commuting copies
of sl(2). This implies that irreducible representations
$[j_1,j_2,j_3]$ of \osp\ are labeled by three spins $j_i$. In
these representations the quadratic Casimir element takes the
value
\begin{equation} \label{Casimir}
C\bigl([j_1,j_2,j_3]\bigr) \ = \ - 4 j_1 (j_1-1) + 2 j_2(j_2+1) + 2 j_3
(j_3+1)\ \ .
\end{equation}
A generic (typical)\footnote{See Appendix~A.} representation possesses dimension
\begin{equation}
D\bigl([j_1,j_2,j_3]\bigr)\  = \ 16 (2j_1+1) (2j_2 + 1) (2j_3 + 1)\ \ .
\end{equation}
The representations of \osp\ that appear in the spectrum of the
Laplacian on the supersphere \Ss\ are not generic. On the
supersphere, wave functions are organized in \osp\ multiplets
$\phi_m, m=0,1,2,\dots$. The first multiplet $\phi_{0}$ consists
of a single function, namely the constant $\phi_{0}=1$. It
transforms in the trivial 1-dimensional representation $[0,0,0]$.
For positive values of $m$, the multiplet $\phi_m$ transforms in
the irreducible representation $[\frac{1}{2},
\frac{m-1}{2},\frac{m-1}{2}]$ of \osp. Consequently, the space
${\cal H}_0$ of square integrable functions on the supersphere
\Ss\ decomposes as follows,
\begin{equation} \label{decH0}
{\cal H}_0 \ \cong \ [0,0,0] \ \oplus\
\bigoplus_{m=1}^\infty\,\left
        [\frac12,\frac{m-1}{2},\frac{m-1}{2}\right]\ \ = \
        \bigoplus_{m=0}^\infty \, \lambda_{m,0}\ .
\end{equation}
Here we have also introduced the symbol $\lambda_{m,0}$ such that
$\lambda_{0,0}$ is the trivial representation and $\lambda_{m+1,0}
= [\frac12,\frac{m}{2},\frac{m}{2}]$. According to eq.\
\eqref{Casimir}, the Laplacian takes the values $\Delta_m = m^2$.
The quadratic dependence on $m$ is similar to the bosonic sphere.
On the other hand, the degeneracies are much larger for the
supersphere. In fact, upon restriction to the bosonic subalgebra,
the eigenspaces of the Laplacian decompose according to
$$
\left.
\left[\frac{1}{2},\frac{k}{2},\frac{k}{2}\right]\right|_{\text{sl(2)}
  \oplus \text{sl(2)}\oplus \text{sl(2)}} \ \cong
\ \left(\frac{1}{2},\frac{k}{2},\frac{k}{2}\right) \oplus
  \left(0,\frac{k+1}{2},\frac{k+1}{2}\right) \oplus
  \left(0,\frac{k-1}{2},\frac{k-1}{2}\right)\ \
$$
for $k = m-1 \geq 1$. When $k=0$, the last term must be omitted.
The formula implies that the dimension $D_k$ of the representation
$\lambda_{k,0}$ is given by $D_k = 4k^2+2$ for $k\geq 1$. This is roughly four
times as large as the dimension of the eigenspaces on the bosonic
sphere $\Sp^3$, as one would expect.
\medskip

It is quite instructive to prove the decomposition \eqref{decH0}.
To this end, let us collect the bosonic coordinate functions $x_i
= :X_i, i=1,\ldots,4$ and the fermionic generators
$\eta_i=X_{4+i}$ into a single multiplet $X$. We recall that the
six functions $X_i$ are subject to the
constraint~\eqref{constraint}. The latter may be recast into the
more covariant form $ X_a X_b J^{ab} = R^2$ by introducing an
appropriate matrix $J =(J^{ab})$. The multiplet $X$ transforms in
the fundamental representation
$\lambda_{1,0}=\left[\frac{1}{2},0,0\right]$ of \osp. When we
restrict from \osp\ to its bosonic subalgebra, $X$ splits into a
4-dimensional multiplet in the $(\frac{1}{2}, \frac{1}{2})$
representation of so(4) $\cong$ sl(2)$\oplus$sl(2) and a
2-dimensional multiplet in the $(\frac{1}{2})$ representation of
sp(2) $\cong$ sl(2). While the former is spanned by the bosonic
coordinate functions $x_i$, the latter consists of the odd
elements $\eta_i$. The algebra $\calH_0$ of functions on \Ss\ is
generated by the six coordinates $X_i$, i.e. every square
integrable function can be arbitrarily well approximated by a
polynomial in $X_i$. The space of polynomials comes with an
integer grading given by the degree of homogeneity. Since the
homogeneous polynomials transform in the graded symmetric tensor
product of the fundamental representation $\lambda_{1,0}$, one
might be inclined to identify the direct sum $S\lambda_{1,0}=
\bigoplus\lambda_{1,0}^{\otimes_s}$ of all graded symmetric tensor
powers of the fundamental representation with the space $\calH_0$.
Such an identification, however, would disregard the defining
equation \eqref{constraint} of the supersphere. The constraint
\eqref{constraint} generates an ideal in the symmetric tensor
algebra $S\lambda_{1,0}$ that has to be divided out in order to
avoid overcounting of states. The two-fold symmetric tensor power
of the fundamental representation, for example, is given by
$\lambda_{1,0}^{\otimes_s 2}=[0,0,0]\oplus \lambda_{2,0}$. The
constraint~\eqref{constraint} identifies the multiplet $[0,0,0]$
with the constant function. The latter has been counted already by
the very first term $\lambda_{1,0}^{\otimes_s 0}= [0,0,0]$.
Consequently, when considering the space of homogeneous
polynomials in $X_i$ up to degree $m$, we have to quotient out the
subspace of polynomials that contain the factor $X_aX_bJ^{ab}$,
which is isomorphic to the space of homogeneous polynomials of
degree less or equal to $m-2$. Thereby we are led to the following
expression for $\calH_0$, \beqa \calH_{0}=\lim_{N\rightarrow
\infty} \left(\bigoplus_{m=0}^N\lambda_{1,0}^{\otimes_s m}\right)
\Big/\left(\bigoplus_{m=0}^{N-2}\lambda_{1,0}^{\otimes_s m}
\right)
 = \bigoplus_{m=0}^{\infty} \lambda_{m,0} \ =\  [0,0,0]\oplus
\bigoplus_{k=0}^{\infty} \left[\frac{1}{2},
\frac{k}{2},\frac{k}{2}\right] \eeqa
where we have used the tensor
product decomposition\footnote{ By $[x]$ we mean the floor function of $x$.} $\lambda_{1,0}^{\otimes_s m}\cong
\bigoplus_{i=0}^{[m/2]} \lambda_{m-2i,0}$ and the identity
$\lambda_{k+1,0}= [\frac{1}{2}, \frac{k}{2}, \frac{k}{2}]$ for
$k\geq 0$.

Before we conclude this subsection, let us briefly construct the
partition function for a particle on the supersphere. By this we
mean the quantity
$$ Z_0 \ = \ Z_0(z_1,z_2,z_3) \ = \ \tr_{\calH_0} (z_1^{H^1} z_2^{H^2}
z_3^{H^3}) $$ where $H^i$ are the three Cartan generators and the
trace is taken evaluated in the space $\calH_0$ of square
integrable functions on the supersphere \Ss. The results we
sketched in the previous paragraphs imply that
\begin{eqnarray}\label{Z0} & & Z_0 \ =\   1 +
\sum_{m=0}^\infty \chi_{[\frac{1}{2},\frac{m}{2},
\frac{m}{2}]}(z_1,z_2,z_3) \\[2mm]
\text{where} & & \chi_{[\frac{1}{2},\frac{m}{2},
\frac{m}{2}]}(z_1,z_2,z_3)\ = \
 \chi_{(\frac{1}{2},\frac{m}{2},\frac{m}{2})} + \chi_{
  (0,\frac{m+1}{2},\frac{m+1}{2})} +
  \chi_{(0,\frac{m-1}{2},\frac{m-1}{2})}\ \ .
\end{eqnarray}
In the second line the last term should be omitted for $m=0$ and
the character $\chi_{(j_1,j_2,j_3)} = \prod_i \chi_{j_i}(z_i)$
denotes a product of bosonic sl(2) characters. The partition
function $Z_0$ can be written in a different form that mimics our
proof of the formula \eqref{decH0}. To this end, let us consider
the module $S\lambda_{1,0}$. We think of it as being generated by
four bosonic coordinates in the $(\frac{1}{2}, \frac{1}{2})$
representation of sl(2)$\oplus$sl(2) $\cong$ so(4) along with the
two fermionic ones in the $(\frac{1}{2})$ representation of sl(2)
$\cong$ sp(2). On $S\lambda_{1,0}$ we introduce the number
operator $N$ that counts the number of bosonic and fermionic
coordinate functions in a given monomial. Since there are no
non-trivial relations in $S\lambda_{1,0}$ we can easily compute
$$
Z^S(t) \, = \, \tr_{S\lambda_{1,0}} (t^N z_1^{H^1} z_2^{H^2}
z_3^{H^3}) \, = \, \frac{(1+z_1^{\frac{1}{2}} t)
(1+z_1^{-\frac{1}{2}} t)}{(1-z_2^{\frac{1}{2}}z_3^{\frac{1}{2}} t)
(1-z_2^{\frac{1}{2}}z_3^{-\frac{1}{2}} t)(1-z_2^{-\frac{1}{2}}
z_3^{\frac{1}{2}} t)(1-z_2^{-\frac{1}{2}}z_3^{-\frac{1}{2}} t)}\ .
$$
Multiplying this quantity with $(1-t^2)$ implements the constraint
\eqref{constraint} on the level of generating functions. We can
then remove $t$ by sending it to $t\rightarrow 1$. The result is a
rather elegant new formula for the partition function $Z_0$,
\begin{equation} \label{Z0alt}
Z_0(z_1,z_2,z_3) \ =\ \lim_{t\rightarrow 1}\left[ (1-t^2)
Z^S(t;z_1,z_2,z_3) \right]\ .
\end{equation}
If the quotient is expanded in a Taylor series and expressions are
reorganized into characters of \osp\ we recover our previous
result \eqref{Z0}.

\subsection{The complete boundary spectrum}

Now let us turn to the spectrum of the PCM \eqref{PCM} at the
special point $R=\infty$ where our field theory becomes free. At
this point, the fields are easy to list and their weights agree
with their classical values. For simplicity, we shall study the
boundary spectrum of a volume filling brane, i.e.\ with Neumann
boundary conditions imposed on all fields of the model. In this
case it suffices to consider the derivative $\partial_u$ along the
boundary, rather than two world-sheet derivatives $\partial$ and
$\bar \partial$. From now on, the letters $x_i = x_i(u), \eta_a=
\eta_a(u)$ and $X_i = X_i(u)$ shall denote boundary fields rather
than coordinate functions.

So, let us begin to analyze the space $\calH$ of boundary fields.
Obviously, ${\cal H}$ is spanned by monomials $\Phi$ of the form
\beqa\label{monomial}  \Phi\ =\
\prod_{i_0}X_{i_0}\prod_{i_1}\partial X_{i_1}\prod_{i_2}\partial^2
X_{i_2}\cdots \ \ . \eeqa The number of factors involving no, one,
two etc. derivatives $\partial =
\partial_u$ of the fundamental fields is arbitrary. Let us stress
at this point already that the defining relation
\eqref{constraint} of the supersphere imposes many relations
between monomials of the form \eqref{monomial}. The space $\calH$,
comes equipped with an integer grading, i.e.\
$\calH=\bigoplus_{n=0}^{\infty} \calH_n$, where $\calH_n$ is
spanned by monomials $\Phi$ with a total number $n$ of
derivatives. The expression $X_a \partial X_b \partial^4 X_c$, for
example, is an element of $\calH_5$.

Associated with the integer grading of the state space ${\cal H}$
there is a corresponding decomposition of the partition function
\begin{equation} \label{Z} Z(q) \ = \ %
\str_{\calH}(q^{L_0-\frac{c}{24}}z_1^{H^1}z_2^{H^2}z_3^{H^3})
  \ =\ %
  q^{-\frac{1}{24}} \, \sum_{n=0}^\infty \, Z_n \, q^n \ .
\end{equation}  The coefficients $Z_n = Z_n(z_i)$ are (infinite) linear
combinations of \osp\ characters. A formula for $Z_0$ was
discussed in the previous subsection. In the present context it
encodes all information on the \osp\ transformation law of fields
with conformal weight $h=0$. These are in one-to-one
correspondence with functions on the supersphere \Ss\ (recall that
we are working at $R =\infty$).

Let us now turn to states involving a single derivative $\partial$.
Since $\calH_1$ is built from fields of the form $\phi_n(X_i)
\partial X_i$, where $\phi_n\in \calH_0$,  one might at first sight
suspect that $Z'_1 = Z_0
\chi_{\lambda_{1,0}}$ coincides with $Z_1$. But this is not true
since it actually counts many fields twice. So far, we have not
 accounted for the derivative of the supersphere relation
\eqref{constraint}. Taking the derivative of this constraint we
find
$$ \sum_{i,j} X_i\partial X_j J^{ij}\ =\ 0\ . $$
This additional condition tells us to subtract $Z_0$ from $Z'_1$.
Hence we find that $Z_1=Z_0
(\chi_{\lambda_{1,0}}-\chi_{\lambda_{0,0}})$ and a simple computer
program can decompose this product into characters of \osp,
leading to
\begin{equation} \label{Z1}
Z_1 \ = \
\sum_{k=0}^{\infty}\left(\chi_{[1,\frac{k}{2},\frac{k}{2}]}+
       \chi_{[\frac{1}{2},\frac{k}{2},\frac{k}{2}]}\right)\ \ .
\end{equation}
In order to gain some more familiarity with the state counting we
invite the reader to construct the contribution $Z_2$ of fields
with two derivatives to the total partition function. The answer
is given by
\begin{eqnarray} \label{Ztwo}
Z_2 \ &=& \ \chi_{[0,0,0]}+2\sum_{k=0}^{\infty}
     \chi_{[\frac{1}{2},\frac{k}{2},\frac{k}{2}]}+
      \chi_{[1,0,0]}\nonumber\\&+&\sum_{k=1}^{\infty}\left(
       \chi_{[1,\frac{k+1}{2},\frac{k-1}{2}]}+
        \chi_{[1,\frac{k-1}{2},\frac{k+1}{2}]}+
         2\chi_{[\frac{1}{2},\frac{k}{2},\frac{k}{2}]}+2\chi_{[1,\frac{k}{2},\frac{k}{2}]}
         \right)\ \ .
\end{eqnarray}
Instead of explaining this formula we shall turn to the higher
subtraces $Z_i$ right away. To begin with, let us enumerate
expressions in which no field appears without derivative and where
the total degree of the derivatives adds up to $n$. There are
$p(n)$ of these terms, where $p(n)$ is the number of partitions of
the integer $n$. We shall denote the set of partitions by $P(n)$
and think of their elements as sequences $\mu = (\mu_i, i=1,2,3,
\dots)$ such that $\sum i \mu_i = n$. With $n=3$, for example, we
have to consider terms involving $\partial^3 X_i$, $\partial^2 X_i
\partial X_j$ and $\partial X_i\partial X_j
\partial X_k$ corresponding to the sequences $(\mu_1,\mu_2,\mu_3)
= (0,0,1),(1,1,0)$ and $(3,0,0)$, respectively. In our notations
we shall suppress the infinite number of zero entries to the right
of the last non-zero one. To each partition $\mu \in P(n)$, we
associate the trace $\chi_{\lambda_{1,0}^{\otimes \mu}}$ over the space
$\lambda_{1,0}^{\otimes_s \mu_1}\otimes \lambda_{1,0}^{\otimes_s
\mu_2} \cdots$,
\begin{equation} \label{chimu}
\chi_{\lambda_{1,0}^{\otimes\mu}} (z_1,z_2,z_3) \ = \ \prod_{i=1}^{\infty} \,
\chi_{\lambda_{1,0}^{\otimes_s \mu_i}} \ \ .
\end{equation}
The factors on the right hand side involve traces over the
$\mu_i^{\text{th}}$ symmetric tensor product of the fundamental
representation $\lambda_{1,0}$. Such factors arise from the
product of $\mu_i$ derivatives of order $i$ of the fundamental
field multiplet. Let us now set $Z_n^{\prime} = Z_0 \sum_{\mu\in
P(n)} \chi_{\lambda_{1,0}^{\otimes\mu}}$ to be $Z_0$ multiplied
with the sum of the $p(n)$ traces \eqref{chimu}. Clearly, $Z'_n$
is not the same as $Z_n$. In fact, we still have to correct for
some overcounting, since we have to subtract all possible
derivatives of degree up to $n$ of the supersphere relations
\eqref{constraint}. Each one of the $p(n)$ partitions  $\mu \in
P(n)$ has to be investigated on its own in order to understand
which relations apply to it. Suppose that for a given partition
$\mu$, the entry $\mu_j$ does not vanish. This means that the
corresponding  fields contain a factor $\partial^{j}X_a$. Hence,
there exist relations between such fields that arise from the
$j^{\text{th}}$ derivative of the supersphere relation
\eqref{constraint}. These must be removed. We may formalize this
prescription by introducing the special partitions $\epsilon^i$
which have a single entry $\epsilon^i_i =1$ in the $i^{\text{th}}$
position and are zero otherwise. The sequence $\epsilon^i$ is an
element of $P(i)$. Let us also denote by $\mu-\epsilon^i$ the
partition from $P(n-i)$ that is obtained by subtracting the
entries. If the resulting sequence contains a negative entry,
i.e.\ if $\mu_i =0$, then we set
$\chi_{\lambda_{1,0}^{\otimes(\mu-\epsilon^i)}}=0$. With these
notations, we can now formalize our resolution for the issue of
overcounting. Taking into account the constraints imposed by the
$i^{\text{th}}$ derivative of \eqref{constraint} amounts to
subtracting from $Z_n'$ all functions of the form
$Z_0\chi_{\lambda_{1,0}^{\otimes(\mu-\epsilon^i)}}$. Here, $\mu
\in P(n)$ and $i$ runs through all integers $i=1,2, \dots$ such
that $\mu_i \neq 0$. After removing all these terms from
$Z_n^{\prime}$ we realize that we actually overdid things with our
correction. In fact we have deleted those expressions for which
two ore more relations are simultaneously fulfilled, so that we
need to put them back in. Thus, we must add all the terms
$Z_0\chi_{\lambda_{1,0}^{\otimes(\mu-\epsilon^i-\epsilon^j)}}$
with $i < j$. The resulting expression overcounts those
polynomials that obey three different relations, etc. A simple
induction leads to the following expression for $Z_n$ \beqa Z_n\
=\ Z_0\, \sum_{\mu\in P(n)}
\left(\chi_{[\frac{1}{2},0,0]^{\otimes\mu}}-\sum_{i=1}^{n}
\chi_{[\frac{1}{2},0,0]^{\otimes(\mu-\epsilon^i})}+\sum_{i<j=1}^{n}
\chi_{[\frac{1}{2},0,0]^{\otimes(\mu-\epsilon^i-\epsilon^j)}}-\cdots\right)
\ . \eeqa All notations that are used in this expression have been
introduced in the preceding paragraph. We have placed the
subscript $\lambda_{1,0} = [\frac12,0,0]$ back on the symbol
$\chi$ to emphasize the relation to the fundamental multiplet. The
reader is invited to check that our general formula for $Z_n$
reproduces the previous expressions (\ref{Z0},\ref{Z1},\ref{Ztwo})
for $Z_n$ when $n \leq 2$.
\medskip

Having found a formula for $Z_n$, we can insert it into our
general prescription \eqref{Z}. The result is, $$  Z\, =\,
q^{-\frac{1}{24}}\ Z_0\ \sum_{n=0}^{\infty}q^n\sum_{\mu\in P(n)}
\left(\chi_{[\frac{1}{2},0,0]^{\otimes\mu}}-\sum_{i=1}^{n}
\chi_{[\frac{1}{2},0,0]^{\otimes(\mu-\epsilon^i)}}+\sum_{i<j=1}^{n}
\chi_{[\frac{1}{2},0,0]^{\otimes(\mu-\epsilon^i-\epsilon^j)}}-\cdots\right)\ .
$$
Now, since $\mu-\epsilon^j$ is a partition in $P(n-j)$, we
are led to the idea of combining in the above alternating sum all
those terms that belong to partitions of the same size. Denoting
by $p_d(x;y)$ the function that counts the number of distinct,
i.e. whose elements are all different, partitions of $x$ with
exactly $y$ elements, we leave to the reader the combinatorial
homework to deduce \beqa & & Z \ =\ q^{-\frac{1}{24}}\ Z_0\
\sum_{n=0}^{\infty} q^n\left(\sum_{j=0}^n
\underbrace{\left(\sum_{k=0}^j(-1)^k p_d(j;k)\right)}_{=:c_j}
\sum_{\mu\in P(n-j)}\chi_{[\frac{1}{2},0,0]^{\otimes\mu}}\right)
\nonumber\\[2mm]
&=&q^{-\frac{1}{24}}Z_0\sum_{n,j=0}^{\infty} q^n c_j \sum_{\mu\in
P(n-j)}\chi_{[\frac{1}{2},0,0]^{\otimes\mu}} \, =\,
q^{-\frac{1}{24}}Z_0\left(\sum_{j=0}^{\infty} c_j q^j\right)
\sum_{n=0}^{\infty} q^n\sum_{\mu\in
P(n)}\chi_{[\frac{1}{2},0,0]^{\otimes\mu}}
\nonumber\\[2mm] &=& q^{-\frac{1}{24}}\ Z_0\ \phi(q)\ \sum_{n=0}^{\infty}
q^n\ \sum_{\mu\in P(n)} \chi_{[\frac{1}{2},0,0]^{\otimes \mu}}\ .
  \label{eq:FullDecomposition}
\eeqa
The numbers $c_j$ can easily be recognized as the coefficients in
the Taylor expansion of the Euler $\phi$-function. In fact the
generating function for distinct partitions of a number $n$ into
precisely $l$ distinct numbers is given by
\begin{equation}
  \prod_{k=1}^{\infty}(1+zq^k)
  \ =\ \sum_{n=0}^\infty\sum_{l=0}^n\,p_d(n;l)\,z^l\,q^n\ \ .
\end{equation}
  For $z=-1$ the left hand side reduces to the Euler function
  $\phi(q)$ while the right hand side gives the sum
  $\sum_{n=0}^{\infty} c_nq^n$. Note that during the resummation in
  the second line of eq.~\eqref{eq:FullDecomposition} we could drop a
  number of terms since $P(n)$ is empty for $n<0$.
The result \eqref{eq:FullDecomposition} has a rather surprising
interpretation. It tells us that we may at first discard all the
derivatives of the supersphere relations for the computation of
subtraces $Z_i$. Derivatives of eq.\ \eqref{constraint} may then
simply be taken into account by multiplying the result with the
Euler function $\phi(q)$.

The conclusion of the previous discussion may now be employed to
derive a much simpler formula for the partition function which
generalizes the expression \eqref{Z0alt} for $Z_0$. Without paying
respect to the supersphere relations, it is straightforward to
enumerate derivative fields. Recall that the four fundamental
bosonic fields carry charges $(0,\pm \frac12,\pm \frac12)$ under
the three Cartan generators $(H^1,H^2,H^3)$. Similarly, the two
fundamental fermionic fields are only charged under the first
Cartan generator $H^1$ such that their charges are $(\pm
\frac12,0,0)$. Hence, the partition function can now be
represented in the form
\begin{equation}
\label{The spectrum} Z\ =\ q^{-\frac{1}{24}}Z_0\, \phi(q)\,
\prod_{n=1}^{\infty}\frac{(1+z_1^{\frac{1}{2}}
q^n)(1+z_1^{-\frac{1}{2}}q^n)}{(1-z_2^{\frac{1}{2}}z_3^{\frac{1}{2}}
q^n)(1-z_2^{\frac{1}{2}}z_3^{-\frac{1}{2}}
q^n)(1-z_2^{-\frac{1}{2}} z_3^{\frac{1}{2}}
q^n)(1-z_2^{-\frac{1}{2}} z_3^{-\frac{1}{2}} q^n)}\ \ .
\end{equation}
The infinite product enumerates all states in the unconstrained
state space. According to our previous discussion, the derivatives of the
supersphere constraints can be implemented through a simple
multiplication with the Euler function $\phi(q)$. Our final formula
for the partition function of a volume filling brane in the PCM at
$R=\infty$ is indeed very simple.

\subsection{Casimir decomposition of the boundary spectrum}

The goal of this section is to expand the partition sum \eqref{Z}
of the volume filling brane in terms of \osp\ characters. To be
more concrete, we would like to derive explicit formulas for the
branching functions $\psi^K_{\Lambda}(q)$ in the decomposition
\begin{equation} \label{Zdecform} Z(q,z_1,z_2,z_3)\ =\ {\sum}_{\Lambda} \
\chi^K_{\Lambda}(z_1,z_2,z_3)\ \psi^K_{\Lambda}(q)\ \ .
\end{equation}
Here, the functions $\chi^K_{\Lambda}(z_1,z_2,z_3)$ are characters
of the Kac modules\footnote{Again, see Appendix A.} $K_{\Lambda}$ of 
\osp. The latter form a basis in the space of all characters so that 
the expansion
coefficients are uniquely determined. Finding an explicit formula
for the branching functions $\psi^K_{\Lambda}(q)$ is the main
result of this section. The final expression will take the
following form
\begin{equation}  \begin{split} \psi^K_{[j_1,j_2,j_3]}(q) & = \
\frac{q^{2j_1(j_1-1)-j_2(j_2+1)-j_3(j_3+1)}}{\eta(q)\phi(q)^3}
\sum_{n,m=0}^{\infty}(-1)^{m+n}q^{\frac{m}{2}(m+4j_1+2n+1)+
\frac{n}{2}+j_1} \\[2mm] & \hspace*{2cm} \times \,
\left(q^{(j_2-\frac{n}{2})^2}-q^{(j_2+\frac{n}{2}+1)^2}\right)
\left(q^{(j_3-\frac{n}{2})^2}-q^{(j_3+\frac{n}{2}+1)^2}\right)\ .
\label{psiK}
\end{split}
\end{equation}
Let us add two remarks here. To begin with, the decomposition 
\eqref{Zdecform} of the supersphere partition function has 
also been considered in the work of Candu and Saleur 
\cite{Candu:2008vw,Candu:2008yw}. In their context, the branching 
functions $\psi^K$ are related to representation spaces of 
the so-called Brauer algebra. The connection has interesting 
implications, but it does not provide explicit formulas for 
$\psi^K$. Our formula \eqref{psiK} has not appeared in the 
literature before.     
In addition, we would want to stress that the decomposition 
of the partition function into characters of {\em Kac modules} 
is a somewhat formal procedure that does not fully capture
the representation content of the spectrum, at least not for the
atypical sector of the theory. One may notice, for example, that
some of the expansion coefficients $C_n$ in $\psi^K_{\Lambda}(q) =
\sum C_n q^n$ are negative. Only for typical $\Lambda$ will the
$c_n = C_n^\Lambda$ are positive. For atypical representations
$\Lambda$, on the other hand, the characters $\chi_{\Lambda}^K$ of
the Kac modules have to be decomposed into characters of
irreducible atypical representations $\chi_{\Lambda}$ as described
in \eqref{Formula for the irreducible characters} in order to
obtain branching functions with non-negative integral
multiplicities.

The proof of eq.\ \eqref{psiK} proceeds in several steps. To begin
with, we shall decompose the partition function into
representations of the bosonic subalgebra of \osp. Our second step
then is to recombine bosonic characters into the characters of
full \osp\ multiplets. Once this is achieved, the resulting
expressions still require some resummation in order to bring them
into a more appealing form.

In our computation, we shall split the full partition function
into three different parts and decompose them separately before
putting all this together. We shall start with the fermionic
contributions in the numerator of the partition function
\eqref{The spectrum}. Apart from the factors that arise from
derivative fields, there are also two terms in $Z_0$ that account
for fermionic zero modes. We may simply set the parameter $t$ to
$t=1$ in those two factors and combine them with the $q$-dependent
terms in the numerators of eq.\ \eqref{The spectrum} to obtain
\beqa Z^{\text{F}}(q,z_1) &:= &  \prod_{n=0}^{\infty}\,
(1+z_1^{\frac{1}{2}} q^n)\, (1+z_1^{-\frac{1}{2}} q^n)\ =\
(1+z_1^{\frac{1}{2}})\, \prod_{n=0}^{\infty} \,
(1+z_1^{\frac{1}{2}} q^{n+1})\,
(1+z_1^{-\frac{1}{2}} q^n)\nonumber\\[2mm] &=& q^{-\frac{1}{8}} \,
\left( z_1^{-\frac{1}{4}} +z_1^{\frac{1}{4}}\right)\,
\frac{1}{\phi(q)}\  \theta_2(z_1^{\frac{1}{2}}|q) \ =\
\frac{1}{\phi(q)} \, \sum_{n\in \mathbb{Z}}\, z_1^{\frac{n}{2}}\,
\left(q^{\frac{n(n+1)}{2}}+q^{\frac{n(n-1)}{2}}\right)
\nonumber\\[2mm] &=& \frac{1}{\phi(q)}\, \sum_{n=0,\frac{1}{2},1,...}
\left(q^{n(2n+1)}+q^{n(2n-1)}-q^{(n+1)(2n+3)}-q^{(n+1)(2n+1)}\right)
\, \chi_n(z_1) \ \ . \nonumber \eeqa Along the way we have used a
number of simple identities\footnote{See equation
\eqref{thetaidentity}.} for $\theta$-functions. As a result,
all the fermionic contributions to the partition function have
been decomposed explicitly into multiplets of the even part of
\osp. Note that the two fermions transform non-trivially only under
the first subalgebra sl(2) and hence there is no dependence on
$z_2$ and $z_3$ this time.

The second piece of the partition function \eqref{The spectrum}
that we would like to split off concerns the bosonic zero modes,
i.e.\ the denominator of the minisuperspace partition function
$Z_0$. Its decomposition into bosonic representations is
straightforward \beqa \lim_{t\rightarrow 1}
\frac{1-t^2}{(1-z_2^{\frac{1}{2}}z_3^{\frac{1}{2}}
t)(1-z_2^{\frac{1}{2}}z_3^{-\frac{1}{2}}
t)(1-z_2^{-\frac{1}{2}}z_3^{\frac{1}{2}}
t)(1-z_2^{-\frac{1}{2}}z_3^{-\frac{1}{2}}
t)}=\sum_{n=0,\frac{1}{2},1,...} \chi_{n}(z_2)\chi_n(z_3)\ \ .
\eeqa
  Note that the sum of characters on the left hand side encodes
the well-known spectrum of a bosonic 3-sphere $\Sp^3 \cong\text{SU}(2)$.
Therefore we can just state this equality without any detailed
calculation. The commuting left and right invariant vector fields
are generated by the second and third copy of sl(2) within the
even part of \osp. Hence, there is no dependence on the parameter
$z_1$.

It remains to analyze the $q$-dependent factors in the denominator
of the partition function~\eqref{The spectrum}. Their contribution
may be expanded as follows \beqa && \hspace*{-1cm}
\prod_{n=1}^{\infty}\left((
1-z_2^{\frac{1}{2}}z_3^{\frac{1}{2}}q^n)
(1-z_2^{\frac{1}{2}}z_3^{-\frac{1}{2}}q^n)
(1-z_2^{-\frac{1}{2}}z_3^{\frac{1}{2}}q^n)
(1-z_2^{-\frac{1}{2}}z_3^{-\frac{1}{2}}q^n)\right)^{-1}
\nonumber\\[2mm]
&=&\left(\sum_{n\in \mathbb{Z}}
\frac{z_2^{\frac{n}{2}}z_3^{\frac{n}{2}}}{\phi(q)^2} \,
\sum_{m=0}^{\infty}(-1)^m
\left(q^{\frac{m}{2}(m+2n+1)}-q^{\frac{m}{2}(m+2n-1)}\right)
\right)\ \times \ \left(\phantom{\sum_{m}^\infty} z_3
\longrightarrow z_3^{-1} \phantom{\sum_m^\infty} \right) \nonumber\\[2mm]
&=&\sum_{\substack{k,l\in \mathbb{Z}\\k+l\in 2\mathbb{Z}}}
\frac{z_2^{\frac{k}{2}}z_3^{\frac{l}{2}}} {\phi(q)^4} \,
\sum_{n,m=1}^{\infty}(-1)^{n+m} q^{k\frac{n+m}{2}+l\frac{n-m}{2}}
\left(q^{\frac{n(n+1)}{2}}-q^{\frac{n(n-1)}{2}}\right)
\left(q^{\frac{m(m+1)}{2}}-q^{\frac{m(m-1)}{2}}\right)\nonumber\\[2mm]
&=&\frac{1}{\phi(q)^4}\, \sum_{\substack{k,l\in \mathbb{N}\\k+l\in
2\mathbb{N}}}\sum_{n,m=1}^{\infty} \, (-1)^{n+m} \, \frac{
(1-q^n)(1-q^m)(1-q^{n+m})(1-q^{n-m})}{ q^{-(k(n+m)+l(n-m)+n(n-1)+
m(m-1))/2}} \ \chi_{\frac{k}{2}}(z_2)\chi_{\frac{l}{2}}(z_3)\ .
\nonumber \eeqa In the first line of the above computation we have
used the lemma \eqref{laurentidentity}. Since all the contributions
being captured by this computation are associated with bosonic
fields, characters with a non-trivial $z_1$ dependence do not
arise.

In order to obtain the decomposition of $Z$ into characters of
$\osp_{\bar{0}}\cong\text{sl}(2)\oplus\text{sl}(2)\oplus\text{sl}(2)$, we need to put the
results from the preceding three computations together into one
expression. The answer contains products of characters which
depend on the same variables $z_2$ and $z_3$. These products can
be re-expanded with the help of the following auxiliary formula
\begin{equation}
\begin{split}
&
\sum_{p=0}^{\infty}\chi_{\frac{p}{2}}(z_2)\chi_{\frac{p}{2}}(z_3)
\sum_{\substack{k,l\in\mathbb{N}\\ k+l\in 2\mathbb{N}}}
a_{k,l}\chi_{\frac{k}{2}}(z_2)\chi_{\frac{l}{2}}(z_3) \\[2mm]
& \hspace*{2cm}  =\ \sum_{\substack{k,l\in \mathbb{N}\\k+l\in
2\mathbb{N}}}
\chi_{\frac{k}{2}}(z_2)\chi_{\frac{l}{2}}(z_3)\left(\sum_{p=0}^{\infty}
\sum_{r=0}^{\min\{k,p\}}\sum_{s=0}^{\min\{l,p\}}
a_{|k-p|+2r,|l-p|+2s}\right) \end{split} \end{equation} which
holds for an arbitrary set of numbers $a_{k,l}$. When applied to
the case at hand, we find \begin{equation}
\begin{split} Z&=\ \frac{1}{\phi(q)^3\eta(q)}\, Z^{\text{F}}(q,z_1) \,
\sum_{\substack{j_2,j_3\in \frac{1}{2}\mathbb{N}\\j_2+j_3\in
\mathbb{N}}}\chi_{j_2}(z_2)\chi_{j_3}(z_3)\sum_{m,n=1}^{\infty}
(-1)^{m+n}q^{\frac{n(n-1)}{2}+\frac{m(m-1)}{2}}\\[2mm]
& \hspace*{4cm} \times\
(1-q^{n+m})(q^{(n-m)(j_2-j_3)}-q^{(n-m)(j_2+j_3+1)})
\end{split} \end{equation}
Thereby, we completed out first task, namely to decompose the full
partition function $Z$ into irreducible representations of the
bosonic subalgebra of \osp.

Our next issue is to combine bosonic characters back into the
characters of Kac modules of \osp. Since the even part of \osp\ is
a subalgebra of \osp, it is clear that the characters of \osp\ Kac
modules, possess a decomposition into characters of the bosonic
subalgebra. These decomposition formulas may be inverted such that
bosonic characters can be written as infinite linear combinations
of \osp\ characters. All necessary details are provided in
Appendix~\ref{ap:Kac}. The resulting expression for the partition function
$Z$ is of the form~\eqref{Zdecform} with \beqa &&\hspace*{-1cm}
\psi^K_{[j_1,j_2,j_3]}(q)\ =\
\frac{1}{\eta(q)\phi(q)^3}\sum_{k=0}^{\infty}
\sum_{m,n=1}^{\infty}\sum_{l=0}^{\infty} (-1)^{m+n+k}
q^{2j_1(j_1+k+2l)}
q^{\frac{n(n-1)}{2}+\frac{m(m-1)}{2}}\nonumber\\[2mm]&&
\hspace*{1.5cm} \times \sum_{r,s=0}^{k}
q^{(n-m)(r-s)}(1-q^{n+m})(q^{(n-m)(j_2-j_3)}-q^{(n-m)(j_2+j_3+1)})
\nonumber\\[2mm]
&&\times\left[q^{j_1+\frac{k+2l}{2}(k+2l+1)}+q^{-j_1+\frac{k+2l}{2}(k+2l-1)}
-q^{5j_1+3+\frac{k+2l}{2}(k+2l+5)}-q^{3j_1+\frac{k+2l}{2}(k+2l+3)}\right]
\nonumber\\[4mm]
&=&\frac{q^{2j_1(j_1-1)}}{\eta(q)\phi(q)^3}\sum_{m,n=1}^{\infty}
\sum_{k=-\infty}^{\infty} (-1)^k\sum_{l=0}^{\infty} q^{j_1(2|k|+4l+1)+
\frac{|k|}{2}(|k|-1)+l(2l+2|k|-1)}(1-q^{|k|+2l+2j_1})\nonumber\\[2mm]
&&\times (-1)^{m+n}q^{\frac{n(n-1)}{2}+\frac{m(m-1)}{2}}q^{(n-m)k}
(1-q^{n+m})(q^{(n-m)(j_2-j_3)}-q^{(n-m)(j_2+j_3+1)})\ . \nonumber
\eeqa We will now make several transformations and resummations in
order to cast this unwieldy expression into the form \eqref{psiK} we
have spelled out above. Making the substitution $n+m=r+2, n-m=s$ with $r\in
\mathbb{N}$ and $s=-r, -r+2, \ldots , r$,  using the trick
\eqref{trick1} and then substituting $r\rightarrow r+1$ gives the
result \beqa &&\hspace*{-5mm}
\psi^K(q)=\frac{q^{2j_1(j_1-1)}}{\eta(q)\phi(q)^3}
\sum_{k=-\infty}^{\infty}\sum_{r,l=0}^{\infty}(-1)^{r+k}
q^{j_1(2|k|+1)+\frac{|k|(|k|-1)}{2}+l(2l+2|k|+4j_1-1)}
\left(q^{|k|+2l+2j_1}-1\right)\nonumber\\[2mm]
&&\!\!\times q^{\frac{(r+2)(r+1)}{2}}\left(q^{(r+1)(j_2-j_3+k)}+
q^{(r+1)(-j_2+j_3-k)}-q^{(r+1)(j_2+j_3+1+k)}-
q^{(r+1)(-j_2-j_3-1-k)}\right).\nonumber \eeqa In order to
simplify the sum over $r$, we now need to split the summation over
$k$ into three parts, according to whether it is positive, zero or
negative. We then recombine the summations over positive and
negative $k$ into a single sum and employ another auxiliary
formula~\eqref{trick2} from Appendix~B to find \beqa
&&\hspace*{-0.5cm} \psi^K_{[j_1,j_2,j_3]}(q)\ =\
q^{2j_1(j_1-1)-j_2(j_2+1)-j_3(j_3+1)} \frac{1}{
{\eta(q)\phi(q)^3}}\ \sum_{l=0}^{\infty}\ \sum_{r=0}^{\infty}\
(-1)^r\ q^{\frac{r}{2}+j_1}\nonumber\\[2mm] &&\times
\left(q^{(j_2-\frac{r}{2})^2}-q^{(j_2+\frac{r}{2}+1)^2}\right)
\left(q^{(j_3-\frac{r}{2})^2}-q^{(j_3+\frac{r}{2}+1)^2}\right)
\Big[q^{l(2l+4j_1-1)}(1+q^{2l+2j_1}) \\[2mm]
&&+\sum_{k=1}^{\infty}(-1)^kq^{j_1(2k+1)+\frac{k(k-1)}{2}
+l(2l+2k+4j_1-1)}(1-q^{r+1})(q^{(r+1)(k-1)}+q^{-(r+1)k})\Big]\ .
\nonumber \eeqa Once again we need to rearrange the sum over $k$.
Terms can be combined into a single summation if we let $l$ run
over half-integers rather than integers. Making the substitutions
$l\rightarrow 2m$ and  $r\rightarrow n$, leads to the formula
\beqa &&\hspace*{-.5cm} \psi^K_{[j_1,j_2,j_3]}(q)\ =\
\frac{q^{2j_1(j_1-1)-j_2(j_2+1)-j_3(j_3+1)}}
{\eta(q)\phi(q)^3}\sum_{n,m=0}^{\infty}\sum_{k=-\infty}^{\infty}
(-1)^{m+n+k}q^{\frac{m}{2}(m+4j_1-1)+\frac{n}{2}+j_1}\nonumber\\[2mm]
&&\times\left(q^{(j_2-\frac{n}{2})^2}-q^{(j_2+\frac{n}{2}+1)^2}\right)
\left(q^{(j_3-\frac{n}{2})^2}-q^{(j_3+\frac{n}{2}+1)^2}\right)
q^{|k|(2j_1+m)+\frac{|k|(|k|-1)}{2}+(n+1)k}\ . \nonumber \eeqa It is
advantageous to split the summation over $k$ again depending on
whether $k$ is negative or non-negative. Then we substitute $r$
for the sum  $r= m+k$ and $s$ for the difference  $s=m-k$. After
some rather trivial but tedious steps we can thereby bring
$\psi^K$ into the form \beqa &&\hspace*{-1cm}
\psi^K_{[j_1,j_2,j_3]}(q)\ =\
\frac{q^{2j_1(j_1-1)-j_2(j_2+1)-j_3(j_3+1)}}{\eta(q)\phi(q)^3}
\sum_{n,m=0}^{\infty}(-1)^{m+n}q^{\frac{m}{2}(m+4j_1+2n+1)+
\frac{n}{2}+j_1}\nonumber\\[2mm] &&\times\
\left(q^{(j_2-\frac{n}{2})^2}-q^{(j_2+\frac{n}{2}+1)^2}\right)\
\left(q^{(j_3-\frac{n}{2})^2}-q^{(j_3+\frac{n}{2}+1)^2}\right)\
\sum_{s=0}^{2m}q^{-s(n+1)}\ . \nonumber \eeqa It is left to the
reader to use lemma \eqref{trivial lemma} in order to show that
this is equal to the formula \eqref{psiK} we spelled out at the
beginning of this section. Before we conclude our discussion of
the large volume limit, let us stress that our decomposition
\eqref{Zdecform} does not imply that states actually transform in
Kac modules of \osp. The partition sum does not contain any
information on how irreducible atypical representations are
actually combined into indecomposables of \osp. For us, the
characters of Kac modules were simply a convenient basis to
use.

\section{The $\mathbf{\text{OSP}(4|2)}$ GN model and the supersphere $\mathbf{\Ss}$}

In this section we shall study the conjectured dual GN model. We
begin with the free bulk theory defined by eq.\ \eqref{FFact}.
After a brief discussion of the bulk spectrum for generic $S$ we
specialize to $S=1$ and re-express the bulk partition function
through characters of the model's affine \aosp\ symmetry at level
$k=1$.\footnote{The discrepancy between our value $k=1$ and the
$k=-1/2$ that appears in the work of Candu and Saleur is entirely
due to different conventions.} In section 3.2 we analyze one
particular symmetry preserving boundary condition and spell out
its spectrum. The latter is then decomposed according to the
action of the global \osp\ symmetry in the third subsection. Once
such a Casimir decomposition has been performed, we can apply the
results of \cite{Quella:2007sg} and determine the boundary
spectrum throughout the entire moduli space that is generated by
the deformation. We shall show that at $R=\infty$ we recover
precisely the spectrum of the volume filling brane in the PCM on
the supersphere \Ss.

\subsection{Free field construction of the bulk theory}
\def\phy{y}

Before we discuss the spectrum and symmetries of the free
Gross-Neveu model \eqref{FFact}, it is useful to recall how things
work for the case $S=0$, i.e. for the fermionic description of the
free boson. As is well known, the compactified free boson at
radius $R=1$ is equivalent to the free field theory of two real
fermions. Each of the two fermionic fields gives rise to a copy of
the Ising model with $c=1/2$. The two factors, however, are
coupled by an orbifold construction to ensure that only sectors
contribute in which both fermions obey the same (anti-)periodic
boundary conditions. In the next few paragraphs we would like to
formalize this construction. It will turn out rather useful for
the generalization to $S>0$.

Let us begin with a few words on the sectors of the critical Ising
model. We recall that the Virasoro algebra with $c=1/2$ possesses
three sectors which we shall label by the conformal weights of
their ground states, i.e.\ through $[0],[1/2]$ and $[\sigma]=
[1/16]$. The character functions of these sectors read as follows,
\begin{equation}
  \label{eq:IsingCharacters}
\chi_\epsilon(q) \ = \ \frac12 \left( \sqrt{\frac{\theta_3}{\eta}}
+ (-1)^{2\epsilon} \sqrt{\frac{\theta_4}{\eta}}\right) \ \ \ \ , \
\ \ \chi_\sigma(q) \ = \ \frac{1}{\sqrt
2}\sqrt{\frac{\theta_2}{\eta}}
\end{equation}
with the slightly unusual notation $\epsilon = 0,1/2$. This will
turn out rather convenient below. The product of two Ising models
contains a special sector $\gamma = [1/2,1/2]$ with weight $h=1$.
It generates an abelian group $\Gamma_0 = \mathbb{Z}_2$ in the
fusion ring. Elements of this group are called simple currents since
their fusion with an arbitrary representation always yields a single
contribution. We claim that the corresponding simple current orbifold
model is equivalent to the compactified free boson at $R=1$.

The construction of a simple current orbifold proceeds in several
simple steps. To begin with, we have to list all sectors $[J]$ of
the theory which possess integer monodromy charge $Q_J(\gamma) =
h_J + h_\gamma - h_{\gamma \times J}$. These are then organized
into orbits ${\cal O}_a$ under the action of the simple current
group $\Gamma$. Each such orbit ${\cal O}_a$ contributes one term
$Z_a$ to the partition function of the orbifold model, with a
coefficient $|\Gamma|/|{\cal O}_a|$ that is given by the ratio
between the order $|\Gamma|$ of the orbifold group and the length
$|{\cal O}_a|$ of the orbit (see e.g.\
\cite{Schellekens:1990xy}). In our case, there exist five sectors
$[J] = [\epsilon_1,\epsilon_2]$ and $[J] = [\sigma,\sigma]$ with
integer monodromy charge. Under the action of $\Gamma_0$ they are
organized into three orbits, two of length two and one that is
left invariant by fusion with $\gamma$. Consequently, the
associated simple current orbifold invariant becomes
\begin{equation} \label{GN0orb}
  Z^{\text{orb}(\Gamma_0)}_{\text{Ising}^2}(q) \ = \
  Z^{{\text{FF}}}_{S=0}(q)
  \ =\ \bigl|\chi_{(0,0)}+\chi_{(1/2,1/2)}\bigr|^2
       +\bigl|\chi_{(0,1/2)}+\chi_{(1/2,0)}\bigr|^2
       +2\bigl|\chi_{(\sigma,\sigma)}\bigr|^2\ \ .
\end{equation}
The characters on the right hand side are products of characters
of the $c=1/2$ Virasoro algebra, i.e.\ $\chi_{(0,1/2)}(q) =
\chi_0(q)\chi_{1/2}(q)$ etc. According to the claims we stated
above, the simple current orbifold \eqref{GN0orb} agrees with the
free boson compactified at radius $R=1$,
\begin{equation}
\label{eq:IsingOrbifold} Z^{{\text{FF}}}_{S=0}(q) \ = \
\frac{1}{|\eta(q)|^2} \sum_{n,w} q^{\frac{1}{8}(n+2w)^2}\bar
q^{\frac{1}{8}(n-2w)^2}\ = \ Z^{R=1}(q) \ .
\end{equation}
The detailed proof of this identity can be found e.g.\ in the
lectures of Ginsparg \cite{Ginsparg:1988ui}. Our aim now is to
extend eq.\ \eqref{eq:IsingOrbifold} to the case $S>0$.
\medskip

For $S>0$, our theory \eqref{FFact} is built from $2S+2$ real
fermions whose properties we have reviewed already. In addition
there are also $S$ free $\beta\gamma$-systems with central charge
$ c = -1$ (see \cite{Lesage:2002ch} for a detailed analysis of
this rather unusual CFT in the context of our work). For \ospn\
symmetry it is necessary that all these fields obey the same
boundary conditions, i.e.\ are either all periodic or all
anti-periodic. Before we spell out the relevant bulk partition
function, we need a bit more background on the
$\beta\gamma$-systems.

As in the case of real fermions, we shall consider sectors which
differ by the choice of boundary conditions on the fields $\beta$
and $\gamma$. Let us introduce a family of ground states
$|\nu\rangle$ for $\nu \in \frac12\mathbb{Z}$. These states are
characterized by the conditions
\begin{equation}
 \beta_{r+\nu} |\nu\rangle\,  =\,  0 \ \ \ , \ \ \ \gamma_{r-\nu}
 |\nu\rangle \, = \, 0 \ \ \ \text{ for } \ \ \ r \, = \,
1/2,3/2,5/2, \dots
\end{equation} From the ground states we generate the corresponding
sectors by application of raising operators. If we assign charges
$q_\beta = 1/2$ and $q_\gamma = -1/2$ to the modes of the fields
$\beta$ and $\gamma$, respectively, and $q_\nu=\nu/2$ to the
ground state $|\nu\rangle $ the generating function for the sector
$\nu$ reads,
\begin{equation} \label{chinu}
\chi^{(\nu)} (q,y) \ = \ q^{\frac{1}{24}-\frac{\nu^2}{2}}\,y^{\frac{\nu}{2}}
\prod_{n=0}^\infty \frac{1}{(1-y^{\frac12}q^{n+\frac12 -
\nu})(1-y^{-\frac12}q^{n+\frac12+\nu})} \ = \ \frac{q^{-\nu^2/2}\,y^{\frac{\nu}{2}}\,
\eta(q)}{\theta_4(q,y^{1/2}q^{-\nu})}
\end{equation}
All the constructed sectors carry an action of an affine
$\widehat{\text{sl}}(2)$ current algebra at level $k=-1/2$. In terms
of the fields $\beta$ and $\gamma$ the three currents are constructed
as follows,
\begin{equation}\label{BC1}
E^1_+(z)
  \ =\ \frac12 \beta^2(z)
  \ ,\quad H^1(z)
  \ =\ -\frac{1}{2}\,(\beta\gamma)(z)
  \ ,\quad E^1_-(z)
  \ =\ -\frac12\,\gamma^2(z)\ \ .
\end{equation}
Consequently, we can decompose the generating functions
\eqref{chinu} into characters of irreducible representations of
$\widehat{\text{sl}}(2)_{-1/2}$. In case of $\chi^{(0)}$, for example,
the decomposition is given by
$$
\chi^{(0)}(q,y)\, =\, \frac{\eta(q)}{\theta_4(q,y^{1/2})}\, = \,
\chi^{k=-1/2}_0(q,y) + \chi^{k=-1/2}_{1/2}(q,y) \ .
$$
The two characters on the right hand side belong to irreducible
highest weight representations with lowest weight $h = \epsilon
\in \{0,1/2\}$,
\begin{equation}
\chi^{k=-1/2}_{\epsilon}(q,y)\ =\ \frac{\eta(q)}{2}
\left[\frac{1}{\theta_4(q,y^{1/2})}+ (-1)^{2\epsilon}
\frac{1}{\theta_3(q,y^{1/2})}\right]\ \ .
\end{equation}
Let us note that the ground states transform in
representations of spin $j = \epsilon$. Nevertheless, we shall
continue to think of the subscript of $\chi$ as the conformal
weight rather than the spin. Similar decomposition formulas exist
for all the other functions \eqref{chinu}. All of them are related
by the action of spectral flow automorphisms. In particular, we
have
\begin{equation}
 \chi^{(1/2)} \ = \ \chi^{k=-1/2}_{\sigma;+} +
\chi^{k=-1/2}_{\sigma;-} \ \ \ \text{with}  \ \
\chi_{\sigma;\pm}(q,y)\ =\ \frac{y^{1/4}\eta(q)}{2}
\left[\frac{1}{i\theta_1(q,y^{-1/2})}\pm
\frac{1}{\theta_2(q,y^{-1/2})}\right] \ \ .
\end{equation}
The two characters on the left hand side belong to the two
irreducible lowest weight representations of the current algebra
with spin $j = 1/4$ and $j=3/4$. Their ground states have the same
conformal weight $h=-1/8$.

We are now ready to discuss the relevant bulk modular invariant
for the theory \eqref{FFact} with $S > 0$. Let us begin with the
product of $S$ $\beta\gamma$-systems and $2S+2$ real fermions.
This theory contains a group $\Gamma_S$ of simple currents that
consists of all elements $\gamma$ of the form
$$
\gamma\, = \, [\epsilon_1, \dots \epsilon_S; \epsilon_{S+1},
\dots, \epsilon_{3S+2}] \ \ \ \text{with} \ \ \epsilon_i \in
\{0,1/2\} \ \ \text{and} \ \ \epsilon \,\equiv\, \sum_{i=1}^{3S+2}
\epsilon_i\,=\,0 \ \text{mod\/}\, 1 \ . $$ The first $S$ entries
of $\gamma$ denote sectors of the $\beta\gamma$-system while the
remaining ones are representing sectors in the Ising models.
Together, the elements $\gamma$ generate the abelian group
$\Gamma_S \cong \mathbb{Z}_2^{3S+1}$.

Let us first deal with the sector involving representations with
vanishing spectral flow, $\nu=0$. Under the action of $\Gamma_S$,
the sectors with vanishing monodromy charge split into two orbits
of maximal length. Hence we are led to the following contribution
to the partition function,
\begin{equation}
  Z^{\text{FF}}_{S,0}(q,y_1,\dots,y_n) \ =\ %
\biggl| {\sum}_{\gamma\in \Gamma_S} \chi_{\gamma \times
[0,\dots,0;0, \dots,0]} \biggr|^2  + \biggl|{\sum}_{\gamma\in
\Gamma_S}
 \chi_{\gamma \times [0,\dots,0;0,\dots,0,1/2]}\biggr|^2\ \ .
\end{equation}
However, the total theory has to be invariant under the spectral
flow symmetry. Hence we have to add twisted contributions
$Z^{\text{FF}}_{S,\nu}$. It was already mentioned above that all
the bosonic ghosts and all the fermions have to have identical
periodicity conditions in order to not to spoil \ospn\ symmetry.
Consequently the spectral flow must act diagonally, i.e.\
simultaneously on all sectors, by half-integer shifts.\footnote{It
is worth mentioning that these diagonal spectral flow
transformations are also the only ones which commute with the
action of the orbifold group. Note also that half-integer spectral
flow on ghosts and fermions implies integer spectral flow on the
currents such as those defined in eq.\ \eqref{BC1} and below.} In
the fermionic factors, spectral flow by $\nu = 1/2$ brings us to
$\sigma$-representations. Integer units of the spectral flow,
however, do not give anything new. In the ghost sectors things
works differently because the application of a diagonal spectral
flow leads to an infinite number of new representations
constructed from the ground states $|\nu\rangle$ for
$\nu\in\frac{1}{2}\Integer$. Since the orbits of the half-integer
spectral flow representations possess a stabilizer subgroup
$\mathcal{S}$ of order $|\mathcal{S}| = 2^{2S+1}$ with respect to
the action of $\Gamma_1$ we finally end up with the partition
function
\begin{equation}
  \begin{split}
    Z^{\text{FF}}_S(q,y_1,\dots,y_S)
    &\ =\ \sum_{\nu\in\frac{1}{2}\Integer}Z^{\text{FF}}_{S,\nu}(q,y_1,\dots,y_S)\\[2mm]
    &\ =\ \sum_{\nu\in\Integer}\Biggl[
          \biggl| {\sum}_{\gamma\in \Gamma_S} \chi^{(\nu)}_{\gamma\times[0,\dots,0;0, \dots,0]}
           \biggr|^2
          + \biggl|{\sum}_{\gamma\in\Gamma_S} \chi^{(\nu)}_{\gamma \times
          [0,\dots,0;0,\dots,0,1/2]}\biggr|^2 \Biggr]\nonumber\\[2mm]
    &\qquad\qquad
 + 2^{2S+1}\sum_{\nu\in\Integer+\frac{1}{2}}\Biggl|\prod_{a=1}^S\chi^{(\nu)}
 (q,y_a)\bigl(\chi_{\sigma}(q)\bigr)^{2S+2}
 \Biggr|^2\ \ . \label{PFFFn}
  \end{split}
\end{equation}
Here, the superscript $(\nu)$ on a function $f(y_i)$ of $S$
variables $y_i$ is defined through the prescription
$f^{(\nu)}(y_i)=q^{-S\nu^2/2}f(y_iq^{-2\nu})$.

The rest of our analysis in this section is now carried out for
the special case of $S=1$. Generalizations to larger values of $S$
shall be differed to the next section. The state space of our
orbifold theory can be equipped with the action of an affine \aosp\
Lie superalgebra. We have already spelled out expressions for the
first set of sl(2) currents in equation \eqref{BC1} above. The
currents associated with the other two copies if sl(2) take the
form
\begin{align} \label{BC2}
  E^2_{\pm}(z)
  &\ =\ \frac{1}{2i}\,\bigl[(\psi_1\psi_3)-(\psi_2\psi_4)\pm i\bigl((\psi_1\psi_4)+
   (\psi_2\psi_3)\bigr)\bigr]\ , \\[2mm]
   H^2(z) &\ =\
   \frac{1}{2i}\,\bigl((\psi_3\psi_4)+(\psi_1\psi_2)\bigr)
  \ \ \ \ , \ \ \ \ \ H^3(z)
  \ =\ \frac{1}{2i}\,\bigl((\psi_3\psi_4)-(\psi_1\psi_2)\bigr)\ ,
 \nonumber \\[2mm]  \label{BC3}
  E^3_{\pm}(z)
  &\ =\ \frac{1}{2i}\,\bigl[(\psi_1\psi_3)+(\psi_2\psi_4)\pm i
  \bigl((\psi_1\psi_4)-(\psi_2\psi_3)\bigr)\bigr]\ \ .
\end{align}
They generate two commuting copies of the current algebra
$\widehat{\text{sl}}(2)_1$. In addition, we can introduce the eight
fermionic currents through the following expressions
\begin{align}
  F^{+++}(z)
  &\ =\ i\beta\left(\psi_3+i\psi_4\right)(z)\ ,&
  F^{+--}(z)
  &\ =\ i\beta\left(\psi_3-i\psi_4\right)(z)\ ,\nonumber\\
F^{++-}(z)
  &\ =\ i\beta\left(\psi_1+i\psi_2\right)(z)\ ,&
  F^{+-+}(z)
  &\ =\ i\beta\left(\psi_1-i\psi_2\right)(z)\ ,\nonumber
\end{align}
and similarly for $F^{-\pm\pm}(z)$ with the field $\beta$ in the
above formulas exchanged with $\gamma$. Note that all terms that
contribute to the seventeen currents are quadratic in the basic
fields. Since by construction these basic fields are either all in
the Neveu-Schwarz sector or in the Ramond sector, the currents
obey periodic boundary conditions on the entire state space. In
order to rewrite
the partition function of our bulk theory in terms of affine \aosp\ %
characters, we recall the following formulas for characters of an
$\widehat{\text{sl}}(2)$ currents algebra at level $k=1$,
$$ \quad \chi_0^{k=1}(q,z) \ = \ \frac{\theta_3(q^2,z)} {\eta(q)}
\ \ \ , \ \ \  \chi_{1/2}^{k=1}(q,z)\ =\ \frac{\theta_2(q^2,z)}
{\eta(q)}\ \ .
$$
The lower index $j = 0, 1/2$ now denotes the spin of
representations of the $\widehat{\text{sl}}(2)$ current algebra. In terms
of characters of the bosonic current algebras, the orbifold
partition function reads
\begin{equation} \label{orbPF}
\begin{split} Z^{\text{FF}}_{S=1}(q,z_i) & \ = \ \sum_{\nu=-\infty}^\infty \,
 \Bigl|\chi^{(\nu)}_{(0;0,0)}(q,z_i)+ \chi^{(\nu)}_{(\frac12;\frac12,\frac12)}
(q,z_i)\Bigr|^2 \ + \\[2mm]
& \hspace*{1cm} +
\, \sum_{\nu= -\infty}^\infty\,  \Bigl|\chi^{(\nu)}_{(0;\frac12,\frac12)}(q,z_i) +
\chi^{(\nu)}_{(\frac12;0,0)}(q,z_i)\Bigr|^2 \
\end{split}
\end{equation}
where the action of the spectral flow involves the first variable
$z_1=y$ only and we have defined
$$ \chi_{(j_1;j_2,j_3)}(q,z_i) \ = \
   \chi^{k=-\frac12}_{j_1}(q,z_1)\  \chi^{k=1}_{j_2} (q,z_2)\
   \chi^{k=1}_{j_3}(q,z_3) \ \ . $$
To compare the formula \eqref{orbPF} with our previous expression
\eqref{PFFFn} one has to specialize to $z_2=z_3=1$. Going one step
further we can combine characters of the bosonic current algebra
into $\wosp$ characters according to,
\begin{eqnarray} \label{B2S1}
\chi_{\{0\}}(q,z_i) & =& \chi_{(0;0,0)}(q,z_i) +
\chi_{(\frac12;\frac12,\frac12)}(q,z_i) \ , \\[2mm] \label{B2S2}
\chi_{\{1/2\}}(q,z_i) & =& \chi_{(0;\frac12,\frac12)}(q,z_i)
+\chi_{(\frac12;0,0)}(q,z_i)\ \ .
\end{eqnarray}
The results of this section may then be summarized through the
following simple formula
\begin{equation} \label{orbPF2}
Z^{\text{FF}}_{S=1}(q,z_i) \ = \ \sum_{\nu=-\infty}^\infty \,
\Bigl|\chi^{(\nu)}_{\{0\}}(q,z_i)\Bigr|^2 +
\sum_{\nu=-\infty}^\infty\,
\Bigl|\chi^{(\nu)}_{\{1/2\}}(q,z_i)\Bigr|^2\ \ ,
\end{equation}
i.e.\ the orbifold partition function is the charge conjugate
modular invariant partition function for the sectors $\{0\}$ and
$\{1/2\}$ of the $\wosp$ current algebra. It is remarkable that
spectral flow relates all the representations occurring here and
that the fusion is purely abelian \cite{Lesage:2002ch}. In contrast to
other WZNW theories on supergroups \cite{Schomerus:2005bf,Saleur:2006tf,Gotz:2006qp,Quella:2007hr} this
guarantees the existence of an ``irreducible'' theory without
logarithmic correlation functions. By fermionizing the $\beta\gamma$
systems and keeping additional zero-modes, however, one can as well
construct a ``logarithmic lift'' of the theory \cite{Lesage:2003kn} (see also
\cite{Saleur:2006tf}).

\subsection{Boundary conditions and their spectra}

In the next step we wish to discuss boundary conditions in the
orbifold theory constructed above. We will focus on a particular
brane. Our choice might seem a bit ad hoc at first, but will later
turn out to be deformed into the space-filling brane of the PCM.
As before, we treat the cases $S=0$ and $S=1$ in some detail and
postpone comments on higher values of $S$ to the following
section.

In the case $S=0$ we need to construct a brane in the orbifold
\eqref{GN0orb} which corresponds to a Neumann brane in the free
boson theory at large radius. But in this case the deformation is
well known. When we reduce the radius from $R=\infty$ to $R=1$ we
pass the self-dual radius where Neumann and Dirichlet branes
cannot be distinguished and get exchanged by T-duality.
Consequently the brane we would like to describe in the free boson
theory at $R=1$ is the Dirichlet brane which has the spectrum
\begin{equation}
  \label{eq:SpectrumZero}
  Z_D^{R=1}(q)
  \ =\ \sum_{w\in\Integer}\frac{q^{\frac{w^2}{2}}}{\eta(q)}
  \ =\ \frac{\theta_3(q)}{\eta(q)} \ \ .
\end{equation}
We will now show how the same spectrum can be obtained from the
orbifold model.

The Ising model is the simplest of the Virasoro minimal models. It
has precisely three different conformal boundary conditions, one
for each of irreducible representations $[0]$, $[1/2]$ and
$[\sigma]=[1/16]$. Here and in the following we shall labels
boundary conditions and sectors by the same symbol. The spectrum
of excitations between any two of these boundary conditions is
described by the respective fusion rules \cite{Cardy:1989ir}. In
order to make contact with the bosonic description, let us try to
rewrite the partition function \eqref{eq:SpectrumZero} through
characters \eqref{eq:IsingCharacters} of the two Ising models.
After simple manipulations we find
\begin{equation}
  \label{eq:SpectrumZeroIsing}
  Z_D^{R=1}(q)
\  =\ \frac{\theta_3(q)}{\eta(q)} \ =\
\chi_{(0,0)}+\chi_{(1/2,1/2)}
       +\chi_{(0,1/2)}+\chi_{(1/2,0)}\ \ .
\end{equation}
The spectrum we find can be considered as the orbit of the sum
$[0,0]\oplus[0,1/2]$ under the action of the orbifold group
$\Gamma_0$. Since $[0,0]\oplus[0,1/2]$ is precisely the fusion
product $[\sigma,0]\times [\sigma,0]$ we conclude that the desired
point-like brane at $R=1$ descends under the orbifold construction
from the boundary condition $[\sigma,0]$ in the product of two
Ising models. The conclusion is fully consistent with the free
fermion construction of the bosonic current $J \sim \psi_1\psi_2$
of the $R=1$ model. In fact, as is well known, the boundary label
$[0,\sigma]$ corresponds to the gluing conditions
\begin{align}
  \psi_1(z)&\ =\ -\bar{\psi}_1(\bar z)\qquad&
  \psi_2(z)&\ =\ \bar{\psi}_2(\bar z)\qquad\qquad(\text{for }z=\bar z)
\end{align}
in the underlying free fermion description. The sign in the gluing
condition for the first fermionic field is associated with the
non-trivial boundary label $[\sigma]$. It implies that the current
$J \sim \psi_1\psi_2$ satisfies Dirichlet boundary conditions $J =
-\bar J$ all along the boundary.

Let us now turn our attention to the case $S=1$. We would like to
focus on a brane which is associated with the twisted gluing
conditions
\begin{equation} \label{glue}
   J^1(z) \ = \ \bar J^1(\bar z) \ \ , \ \
   J^2(z) \ = \ \bar J^3(\bar z) \ \ , \ \
   J^3(z) \ = \ \bar J^2(\bar z)
\end{equation}
for the bosonic currents $J^i = E^i_a t^a$ all along the boundary
at $z=\bar{z}$. The underlying gluing automorphism $\Omega$
permutes the second and third copy of sl(2) in the bosonic
subalgebra. It can easily be seen that $\Omega$ extends to an
involution on the entire superalgebra \osp. The corresponding
gluing conditions for fermionic currents read,
\begin{align}
\label{glueF}
  F^{\xi\pm\pm}(z)&\ =\ \bar F^{\xi\pm\pm}(\bar{z})&
  F^{\xi\pm\mp}(z)&\ =\ \bar F^{\xi\mp\pm}(\bar{z})\ \ .
\end{align}
A quick look back at the free field realization of the currents
\eqref{BC2} suggests to implement the boundary conditions
\eqref{glue} and \eqref{glueF} through the following gluing
prescription for the fundamental field multiplet,
\begin{align}
  \label{glueFree}
  \psi_1(z)&\ =\ -\bar{\psi}_1(\bar z)\,,&
  \psi_i(z)&\ =\ \bar{\psi}_i(\bar z)\quad(i\neq1)\,,&
  \beta_a(z)&\ =\ \bar{\beta}_a(\bar z)\,,&
  \gamma_a(z)&\ =\ \bar{\gamma}_a(\bar z)\ \ .
\end{align}
Indeed, equations \eqref{glueFree} reproduce the permutation of
currents displayed in eqs.~\eqref{glue} and \eqref{glueF} upon
insertion into eqs.\ \eqref{BC2}.

Just as in the case $S=0$ above, having a non-trivial gluing
condition for the fermion is associated with the occurrence of the
brane label $\sigma$ in the Ising model description. Hence we
propose that the desired orbifold brane may be constructed from
the brane $B = [0,0;\sigma,0,0,0]$ in the covering theory. The
spectrum for the latter is again given by fusion, and taking the
orbit with respect to the orbifold group $\Gamma_1$ one easily
arrives at
\begin{equation}
  Z_{B;S=1}^{\text{FF}}\ =\ \sum_{\gamma\in\Gamma_1}\bigl[
  \chi_{\gamma\times[0,0;0,0,0,0]}+\chi_{\gamma
  \times[0,0;0,1/2,0,0]}\bigr]\ \ .
\end{equation}
For later convenience this result may also be rewritten in terms of
irreducible characters of the underlying bosonic current algebra,
leading to
\begin{equation}
  Z^{\text{FF}}_{B;S=1}(q,z_i)
  \ = \ \chi_{(0;0,0)} + \chi_{(0;\frac12,\frac12)}
        + \chi_{(\frac12;\frac12,\frac12)} + \chi_{(\frac12;0,0)}
  \ = \ \chi_{\{0\}} + \chi_{\{1/2\}}\ \ .
\label{tZ2}
\end{equation}
In the second step we have combined characters of the bosonic
subalgebra into characters of the full $\wosp$, using the formulas
\eqref{B2S1} and \eqref{B2S2}. The spectrum of the orbifold brane
preserves the affine Lie superalgebra, as desired. We also note
that our partition function $Z^{\text{FF}}_{B;S=1}(q)$ is identical to
the one that appeared in the work of Candu and Saleur
\cite{Candu:2008vw,Candu:2008yw}. We shall now see that it is
related through a deformation to the partition function of the
volume filling brane in the PCM model.

\subsection{Casimir decomposition in the free GN model}

Having found the full spectrum of an \osp\ symmetric brane in the
free field theory \eqref{FFact}, our next task is to expand it in
terms of the characters $\chi_\lambda^K$. In other words, we need
to find the branching functions $\psi^K_\Lambda(q)$ in the
decomposition,
 \beqa \label{tZdecform} \tilde Z\ =\
 Z^{\text{FF}}_{B;S=1}(q,z_i)\ = \
 {\sum}_{\Lambda}\ \chi_{\Lambda}^K(z_1,z_2,z_3)
\ \tilde \psi^K_{\Lambda}(q)\ \ .
 \eeqa
 This expansion is of the same
form \eqref{Zdecform} as in the PCM at $R=\infty$. Only the
branching functions $\tilde \psi^K$ are different. The following
short analysis will show that they read
\begin{equation} \label{tpsiK}
\begin{split} \tilde \psi^K_{[j_1,j_2,j_3]}(q) &\  = \
\frac{1}{\eta(q)\phi^3(q)}\sum_{n,m=0}^{\infty}(-1)^{n+m}
q^{\frac{m}{2}(m+4j_1+2n+1)+j_1+\frac{n}{2}}
\\[2mm]& \hspace*{1cm}\times\
(q^{(j_2-\frac{n}{2})^2}-q^{(j_2+\frac{n}{2}+1)^2})
(q^{(j_3-\frac{n}{2})^2}-q^{(j_3+\frac{n}{2}+1)^2})\ \ . \end{split}
\end{equation} Before we derive this formula, we wish to comment on its
implications. A short look back to formula \eqref{psiK} reveals a
remarkable similarity between the two branching functions of the
partition functions $Z$ of the PCM at $R=\infty$ and $\tilde Z$ of
the free fields theory \eqref{FFact}. In fact, they are identical
up to an overall prefactor,
\begin{equation}\label{main}
\psi^K_{[j_1,j_2,j_3]}(q) \ = \
q^{2j_1(j_1-1)-j_2(j_2+1)-j_3(j_3+1)} \
\tilde\psi^K_{[j_1,j_2,j_3]}(q) \ \ . \end{equation} For the time
being this equation may simply be considered a curious observation
regarding the similarities of the two Casimir decompositions. We shall
explain in the next subsection how it relates to the claim that
the boundary spectrum for the PCM at $R=\infty$ may be obtained by
the current-current perturbation \eqref{int} from the free field
theory \eqref{FFact}.

In order to calculate the branching functions $\tilde \psi^K$ from
the partition function $\tilde Z$, we proceed as in section 2.3.
In a first step we shall expand $\tilde Z$ in terms of characters
of the bosonic subalgebra $\osp_{\bar{0}}$. Then we combine the
bosonic building blocks into characters of Kac modules for \osp.
The resulting expression for the branching function will require
only very little additional analysis in order to cast them into
the form \eqref{tpsiK}.

The decomposition of $\tilde Z$ into bosonic characters departs
from the representation \eqref{tZ2} of $\tilde Z$ and then employs
the following expansion formulas for $\widehat{\text{sl}}(2)$
characters into sums of characters of sl(2), \beqa
\label{Decomposition1}
\chi_a^{k=-\frac{1}{2}}(\tau,u)&=&\frac{q^{\frac{1}{24}}}{\phi(q)^2}
\, \sum_{k\in \mathbb{N}+a}
\chi_{k}(z)\sum_{m=0}^{\infty}
(-1)^mq^{\frac{m}{2}(m+4k+1)+k}\left(1-q^{2m+1}\right)\\[2mm]
\label{Decomposition2} \chi_a^{k=1}(\tau,u)&=& \frac{1}{\eta(q)} \
\sum_{m\in \mathbb{N}+a}\chi_m(z)\
\left(q^{m^2}-q^{(m+1)^2}\right)
\eeqa where $a\in \left\{0,\frac{1}{2}\right\}$.
>From the equality \eqref{tZ2} and the two decomposition
formulas \eqref{Decomposition1} and \eqref{Decomposition2} it is
clear that $\tilde Z$ can be written as \beqa \tilde Z\ =\
\sum_{\substack{(j_1,j_2,j_3)\in
\frac{1}{2}\mathbb{N}^3\\j_2+j_3\in \mathbb{N}}}
\chi_{(j_1,j_2,j_3)}(z_1,z_2,z_3)\ \tilde
\psi^B_{(j_1,j_2,j_3)}(q) \eeqa where $\chi_{(j_1,j_2,j_3)}$ are
the characters of the irreducible representations of
$\osp_{\bar{0}}$, as before, and the branching functions
$\tilde \psi^B$ are given by \begin{equation}
 \begin{split}\tilde
\psi^B_{(j_1,j_2,j_3)}(q)\ & =\ \frac{1}{\eta(q)\phi^3(q)}\
\sum_{m=0}^{\infty}\, (-1)^m\, q^{\frac{m}{2}(m+4j_1+1)+j_1}
\, (1-q^{2m+1})\\[2mm]& \hspace*{2.5cm} \times\
(q^{j_2^2}-q^{(j_2+1)^2})\  (q^{j_3^2}-q^{(j_3+1)^2})\ .
\end{split} \end{equation}
Before we proceed let us note that the branching functions $\tilde
\psi^B_\Lambda$ possess the following important symmetry
properties necessary for a proof in Appendix C, \beqa \tilde
\psi^B_{(j_1,j_2,j_3)}(q)\, =\, -\tilde
\psi^B_{(-j_1-1,j_2,j_3)}(q)\, =\, -\tilde
\psi^B_{(j_1,-j_2-1,j_3)}(q)\, =\, -\tilde
\psi^B_{(j_1,j_2,-j_3-1)}(q)\  . \eeqa These imply in particular
that $\psi^B_{(j_1,j_2,j_3)}(q)$ vanishes identically if any of
the spin labels $j_a$ is equal to $j_a = -1/2$. As in our analysis
of the PCM's partition function $Z$ in section 2.3, we can express
all characters of representations of the bosonic subalgebra as
infinite linear combinations of the characters of Kac modules. The
required formulas can be found in Appendix~\ref{ap:Kac}. With
their help we now arrive at the following result for
$\tilde\psi^K_{\Lambda}$, \beqa
\tilde\psi^K_{[j_1,j_2,j_3]}(q)&=&\frac{1}{\eta(q)\phi^3(q)}
\sum_{n,m=0}^{\infty}(-1)^{n+m}\, q^{\frac{m}{2}(m+4j_1+1)+
j_1+mn+\frac{n}{2}}\, (1-q^{2m+1})\nonumber\\[2mm]&&\times
\, \sum_{k=0}^{[\frac{n}{2}]}(q^{(j_2-\frac{n}{2}+k)^2}-
q^{(j_2+\frac{n}{2}-k+1)^2})\ (q^{(j_3-\frac{n}{2}+k)^2}-
q^{(j_3+\frac{n}{2}-k+1)^2})\nonumber\\[2mm]
&=&\frac{1}{\eta(q)\phi^3(q)}\sum_{n,m=0}^{\infty}(-1)^{n+m} \,
q^{\frac{m}{2}(m+4j_1+2n+1)+j_1+\frac{n}{2}}\, (1-q^{2m+1})
\nonumber\\[2mm]&&\times
(q^{(j_2-\frac{n}{2})^2}-q^{(j_2+\frac{n}{2}+1)^2})\
(q^{(j_3-\frac{n}{2})^2}-q^{(j_3+\frac{n}{2}+1)^2})\,
\sum_{k=0}^{\infty}q^{(2m+1)k}\ \ .\nonumber \eeqa The sum over
$k$ at the end of this formula is a simple geometric series which
cancels the last term in the first line. Thereby, we recover the
expression \eqref{tpsiK} we spelled out at the beginning of this
subsection.

\subsection{Deformation from free GN model to free PCM}

The main result of our analysis so far was summarized concisely in
eq.\ \eqref{main}. In order to fully appreciate its content, let
us review a few results from \cite{Quella:2007sg}. In that paper,
the deformation of conformal weights was studied for the WZNW
model on PSL(2$|$2). Many of the central results of
\cite{Quella:2007sg}, however, hold much more generally for models
whose symmetries are described by an affine Lie superalgebra with
vanishing dual Coxeter number.

To begin with, let us specify the bulk perturbation we would like
to consider. As we shall argue momentarily, it is generated by the
field,
\begin{equation} \label{Phi}\Phi \ = \ \sum \kappa_{\mu\nu}
J^\mu(z) \Omega (\bar J^\nu(\bar z)) \  \end{equation} where the
summation extends over all 17 bosonic and fermionic directions.
The automorphism $\Omega$ we inserted here is the same as the
gluing automorphism that was defined implicitly through our gluing
conditions \eqref{glue} and \eqref{glueF} in section 3.2. Note
that the perturbing operator $\Phi$ breaks the global symmetry
from $\osp\otimes\osp$ of the free GN model \eqref{FFact} to the
twisted diagonal subalgebra. In other words, the symmetry
transformations of the perturbed model are generated by elements
of the form $X \otimes 1 + 1 \otimes \Omega(X)$. This means that
any perturbing operator of the form $\Phi$ preserves half of the
global bulk symmetries. What depends on the choice of the
automorphism $\Omega$ is the precise set of transformations that
is preserved. Similar statements can be made about boundary
conditions. As we discussed in section 3.2, the boundary theory we
put forward to compare with the boundary spectrum of the PCM
required to select a non-trivial gluing automorphism $\Omega$. If
this gluing automorphism would differ from the automorphism
$\Omega$ in the definition of $\Phi$, then the boundary condition
and the deformation would preserve different sets of symmetry
generators. Hence, the deformed boundary theory would no longer
possess a global \osp\ symmetry. Such a theory could be conformal,
but it cannot be equivalent to the boundary PCM. Therefore, we
know that the perturbing operator $\Phi$ must involve the same
automorphism $\Omega$ that appeared in the gluing condition for
currents at the boundary. An explicit formula for the operator
$\Phi$ in terms of free fields is derived at the end of appendix
D. The resulting expression agrees with the formula for
$\Sa^{\text{int}}$ we anticipated in the introduction.

Having specified the deforming operator, we are now ready to
discuss the properties of the deformation it generates. Here we
shall closely follow the the recent analysis in
\cite{Quella:2007sg}. Everything we shall claim below is based on
a rather simple mathematical result that was first formulated and
exploited in the work of Bershadsky et.\ al.\
\cite{Bershadsky:1999hk} for psl(N$|$N), but holds equally for
\ospn. Consider some \ospn\ invariant $\Delta$, such as e.g.\ a
conformal weight, and suppose that $\Delta$ may be written as
$\Delta = C_{abc} f^{abc}$ where $f^{abc}$ are the structure
constants of \ospn\ and $C_{abc}$ are some numbers. Then $\Delta$
can be shown to vanish.

We would like to apply this mathematical lemma to the computation
of conformal weights. To evaluate the change of conformal weights
away from the free GN model, we perform a perturbative analysis of
2-point functions in our theory. In any such computation of
perturbed correlators, the initial step is to remove all the
current insertions through current algebra Ward identities. In the
process, pairs of currents get contracted using
\begin{equation} \label{eq:ob1}
J^\mu(z) \, J^\nu(w) \ = \ \frac{i{f^{\mu\nu}}_\sigma}{z-w} \,
J^\sigma(w) + \frac{k\kappa^{\mu\nu}}{(z-w)^2} + \dots \ \sim \
\frac{k\kappa^{\mu\nu}}{(z-w)^2}\ \ .
\end{equation}
The first equality is the usual operator product for \osp\
currents. Since we are only interested in computing the invariants
$h$, we can drop all terms that involve the structure constants
$f$ of the Lie superalgebra \osp. This applies to the first term
in the above operator product which distinguishes the non-abelian
currents from the abelian algebra of flat target spaces. Here and
in the following we shall use the symbol $\sim$ to mark equalities
that are true up to terms involving structure constants. In
conclusion, we have seen that, as far as the computation of
conformal dimensions is concerned, we may neglect the non-abelian
nature of the currents $J^\mu$. Obviously, this leads to drastic
simplifications of the perturbative expansion.

In \cite{Quella:2007sg} several other statements were needed to
study  a deformation that preserved simultaneously both left and
right global symmetries. The perturbation \eqref{int} we consider
here, however, is of a much simpler type. We can therefore
directly move on to evaluate the conformal dimension of boundary
fields. Unlike in \cite{Quella:2007sg}, the following arguments
apply to all boundary conditions, as long as they preserve the
affine \aosp\ symmetry. It does not require any further
assumptions on the localization of the brane. Let $\Psi$ be some
multiplet of boundary fields transforming in a representation
$\Lambda$ of \osp. We denote by $h_0(\Psi)$ the conformal weight
of $\Psi$ at the WZ-point. Upon deformation with the field
\eqref{Phi}, the weight of $\Psi$ behaves as
\begin{equation}
h(\Psi) \ = \ h_0(\Psi)  - \frac12 \frac{g^2}{1+g^2}\ C_2(\Lambda)
\ = \ h_0(\Psi) + \frac12 \left(\frac{1}{R^2}-1\right) \,
C_2(\Lambda)
\end{equation} where $C_2$ is the quadratic Casimir element of the
Lie superalgebra \osp, as before.

Through the Casimir decomposition \eqref{tZdecform} of the boundary partition
function $\tilde Z$ we have separated all boundary fields according
to their \osp\ transformation law. This now allows us to evaluate
the shift of conformal weights for entire blocks rather than
individual field multiplets. More concretely, the conformal
weights of all fields that are counted by the branching function
$\tilde\psi^K_{[j_1,j_2,j_3]}$ undergo the same shift by\footnote{Let
  us recall that all irreducible multiplets that can be tied together in an
  indecomposable representation must have identical Casimir
  eigenvalues, see appendix~A.}
$$ \delta_g(h) \, = \, -\frac12\frac{g^2}{1+g^2} \, C_2[j_1,j_2,j_3]
\ = \ \frac{g^2}{1+g^2}\, \bigl(2j_1(j_1-1) - j_2(j_2+1) -
j_3(j_3+1)\bigr)\ \
$$ upon perturbation with $\Phi$. Thereby, we can spell out the
boundary spectrum of the perturbed model for any choice of $g^2 =
R^2-1$,
\begin{equation}
\begin{split} \label{PFZR}
 \tilde Z_R(q,z_i) & = \ q^{-\frac{1}{24}}\
 {\sum}_{j_i}\ \chi_{[j_1,j_2,j_3]}^K(z_1,z_2,z_3)\  \times \\[2mm]
& \ \ \ \times \ q^{\left(1-\frac{1}{R^2}\right) \left(2j_1(j_1-1)
- j_2(j_2+1) - j_3(j_3+1)\right)}\tilde \psi^K_{[j_1,j_2,j_3]}(q)\
.
\end{split}
\end{equation}
For irrational values of the parameter $R$, the boundary spectrum
is rather rich, containing irrational conformal weights. But as we
reach the special value $R=\infty$, all conformal weights become
integers. Equation \eqref{main} tells us even more: At this
particular point, the perturbed boundary partition function
coincides with the partition function $Z$ of volume filling branes
in the PCM on the supersphere \Ss\ in the limit $R\to\infty$. For 
a few selected multiplets, the deformation from $R=\infty$ to 
$R=1$ had been carried out in \cite{Candu:2008yw}. By performing 
the Casimir decompositions explicitly, we were able to extend such 
studies to the entire spectrum.

\section{Generalization for higher-dimensional superspheres}

The aim of this section is to outline how the previous analysis
may be extended to higher dimensional superspheres. We shall
provide explicit formulas for the relevant boundary spectra of the
PCM at $R=\infty$ and for the free field theory \eqref{FFact}. The
latter are expressed in terms of characters of the affine \aospn\
superalgebra at $k=1$. Note that the level does not depend on $S$.
Since we have not attempted to construct the branching functions
$\psi_\Lambda$ and $\tilde \psi_\Lambda$ for the decomposition
with respect to the global \ospn\ symmetry, we shall content
ourselves with a few non-trivial tests. These are discussed in the
second subsection. We believe that a full analysis, as in the case
of $S=1$, is possible but cumbersome.

\subsection{Partition functions for superspheres at $R=1,\infty$}

The first task is to spell out the spectrum of the PCM with
Neumann boundary conditions at $R=\infty$. It turns out that our
formula \eqref{The spectrum} for $S=1$ admits the following
straightforward generalization, \beqa \label{The spectrumn}
Z^{\text{PCM}}_{N;S} \ =\
q^{-\frac{1}{24}}Z^{(S)}_0\phi(q)\prod_{n=1}^{\infty}
\frac{\prod_{m=1}^{S}(1+y_m
q^n)(1+y^{-1}_mq^n)}{\prod_{k=1}^{S+1}(1-x_kq^n)(1-x^{-1}_kq^n)}\
. \eeqa Here, the subscript $N$ stands for Neumann boundary
conditions and the minisuperspace contribution is given by \beqa
Z^{(S)}_0\ =\ \lim_{t\rightarrow
1}(1-t^2)\frac{\prod_{m=1}^{S}(1+y_m
t)(1+y^{-1}_mt)}{\prod_{k=1}^{S+1}(1-x_k t)(1-x^{-1}_kt)}\ . \eeqa
As before, the factor $Z^{(S)}_0$ describes the space of functions
on \Ssn. As mentioned above, we have not performed the analysis of
section 2.3 for the more general partition function
$Z^{\text{PCM}}_{N;S}$, though this would surely be possible.

Next let us turn to the free GN model \eqref{FFact}. Large parts
of our analysis of the bulk spectrum were already performed for
generic $S$. Once more, the theory possesses an affine \aospn\
symmetry with level $k=1$ (see appendix D for an explicit
construction of the generators in terms of the basic fields). The
bulk theory can be shown to possess a symmetry preserving boundary
condition whose spectrum closely resembles eq.\ \eqref{tZ2}.
Before we are able to spell out the details, we shall quote from
\cite{FrancescoCFT} the following expressions for characters of
the affine Lie algebra $\widehat{\text{so}}(2S+2)$ at level $k=1$,
\begin{equation}
  \begin{split}
\chi^{\text{so}}_{(0)}(q,x_i)
&\ =\ \frac{1}{2\eta(q)^{S+1}}\left(\prod_{i=1}^{S+1}
\theta_3(q,x_i)+\prod_{i=1}^{S+1}
\theta_4(q,x_i)\right)\ ,\\[2mm]
\chi^{\text{so}}_{(\fu)}(q,x_i)&\ =\ \frac{1}{2\eta(q)^{S+1}}\left(\prod_{i=1}^{S+1}
\theta_3(q,x_i)-\prod_{i=1}^{S+1} \theta_4(q,x_i)\right)\ .
\end{split}
\end{equation}
Note that $\widehat{\text{so}}(2S+2)_1$ is part of the bosonic subalgebra
of $\wospn$.  Similarly, we also need the corresponding characters
of the affine $\widehat{\text{sp}}(2S)$ at $k=-\frac{1}{2}$
\begin{equation}
\begin{split}
\chi^{\text{sp}}_{(0)}(q,y_i)&\ =\ \frac{\eta(q)^S}{2}
\left(\frac{1}{\prod_{i=1}^S\theta_4(q,y_i)}+
\frac{1}{\prod_{i=1}^S\theta_3(q,y_i)}\right)\ ,\\[2mm]
\chi^{\text{sp}}_{(\fu)}(q,y_i)&\ =\ \frac{\eta(q)^S}{2}
\left(\frac{1}{\prod_{i=1}^S\theta_4(q,y_i)}-
\frac{1}{\prod_{i=1}^S\theta_3(q,y_i)}\right) \ .
\end{split}
\end{equation}
The characters we have just listed, furnish the basic building
blocks for the relevant characters of our superalgebra $\wospn$ at
level $k=1$,
\begin{equation}
\begin{split}
 \chi^{\text{osp}}_{\{0\}}&\ =\ \chi^{\text{so}}_{(0)}
\chi^{\text{sp}}_{(0)}+\chi^{\text{so}}_{(\fu)}
\chi^{\text{sp}}_{(\fu)}\ ,\\[2mm]
\chi^{\text{osp}}_{\{\fu\}}&\ =\ \chi^{\text{so}}_{(\fu)}
\chi^{\text{sp}}_{(0)}+\chi^{\text{so}}_{(0)}
\chi^{\text{sp}}_{(\fu)}\ \ .
\end{split}
\end{equation}
For a particular choice of
boundary conditions in the free field theory \eqref{FFact} the
boundary partition function takes the following form \beqa
\label{tZ2n} Z^{\text{FF}}_{B;S}(q,z_i)\ =\
\chi^{\text{osp}}_{\{0\}}+
\chi^{\text{osp}}_{\{\fu\}}=\frac{1}{\eta(q)}
\frac{\prod_{i=1}^{S+1}\theta_3(q,x_i)}
{\prod_{j=1}^S\theta_4(q,y_j)}\ ,  \eeqa where the first $S$
variables $z_i = y_i$ are associated with the symplectic part
while the remaining $S+1$ variables $z_{S+i} = x_{i}$ are
affiliated with Cartan elements of the orthogonal subalgebra. Eq.\
\eqref{tZ2n} generalizes equation \eqref{tZ2} to $S \geq 1$.

\subsection{Test of the duality}

As in the previous section, we would like to show that the two
partition functions \eqref{The spectrumn} and \eqref{tZ2n} are
related to each other by deformation with the interaction term
\eqref{int} or, equivalently, by deforming the radius $R$ of the
PCM from $R=\infty$ all the way down to $R=1$. In principle, this
may be achieved by repeating our analysis in sections 2.3 and 3.3
above. The first step is to decompose the partition function
\eqref{tZ2n} of the PCM at $R=\infty$ in terms of character
functions for the global \ospn\ symmetry, \beqa \label{Rinfty}
Z^{\text{PCM}}_{N,S}\ =\ \sum_{\Lambda\in \cJ}
\chi_{\Lambda}^{\text{osp}(2S+2|2S)}(z_i)\, \psi^{(S)}_\Lambda(q)\
, \eeqa where $\cJ$ is the set of all integral dominant labels of
\ospn\ that are compatible with the consistency conditions of
\cite{Frappat:1996pb}. The existence of such a decomposition is
guaranteed, but in case of $S > 1$ explicit formulas for the
branching functions $\psi$ would still need to be worked out.

The second step is to pass from $R=\infty$ to finite values of the
radius. Since all the general results we outlined in section 3.4
hold for any value of $S$, the boundary partition function of the
PCM  at radius $R$ reads \beqa  \label{ZR} Z(R)\ = \
\sum_{\Lambda\in
\cJ}\chi_{\Lambda}^{\text{osp}(2S+2|2S)}(x_i,y_j)\,
\psi^{(S)}_\Lambda(q)\, q^{\frac12 \frac{1}{R^2} \, C({\Lambda})}\
.  \eeqa Here we expressed the partition function through the
branching functions $\psi$ at $R=\infty$ rather than through the
ones at $R=1$, as in section 3.4. Therefore, the coefficient of
the Casimir element had to be properly adjusted. Note also that we
normalized the quadratic Casimir operator such that $C_2(\fu)=1$
for all values of $S$.

For the PCMs on odd dimensional superspheres \Ssn\ to be dual to
the GN model, we would have to find
 \beqa \label{main2}
 Z(R=1)\ =\ Z^{\text{FF}}_{B;S} \ ,  \eeqa provided we have correctly
identified the appropriate boundary condition in the free field
theory \eqref{FFact}. Throughout the last sections, we have
checked relation \eqref{main2} explicitly for $S=1$. It is quite
amusing to verify it also in the much simpler case of $S=0$. When
$S=0$, the decomposition of the partition function at $R=\infty$
into characters of osp(2$|$0)$\cong$ so(2), takes a particularly
simple form, \beqa
Z^{\text{PCM}}_{N,S=0}&=&q^{-\frac{1}{24}}\phi(q)\sum_{n\in
\mathbb{Z}} z^n\sum_{k\in
\mathbb{Z}}\frac{z^k}{\phi(q)^2}\sum_{m=0}^{\infty}
(-1)^m\left(q^{\frac{m+1}{2}(m+2|k|)}-q^{\frac{m+1}{2}(m+2(|k|+1))}
\right)\nonumber\\[2mm]
&=&\frac{1}{\eta(q)}\sum_{n\in \mathbb{Z}} z^n \ = \ \sum_{n \in
\mathbb{Z}} \chi_n(z) \psi^{(0)}_n(q) \ \ , \eeqa with $\chi_n(z)
= z^n$ and $\psi^{(0)}_n(q) = 1/\eta(q)$. Following our equation
\eqref{ZR}, the partition function for radius $R$ becomes
$$Z(R)\ =\ \frac{1}{\eta(q)}\sum_{n\in \mathbb{Z}} \, z^n\,
q^{\frac12 \frac{1}{R^2}n^2}\ . $$ Therefore, at $R=1$ we obtain
\beqa Z(R=1) \ = \ \frac{1}{\eta(q)}\sum_{n\in \mathbb{Z}}z^n
q^{\frac{n^2}{2}} \ = \ \frac{1}{\eta(q)}\sum_{n\in \mathbb{Z}}z^n
q^{\frac{n^2}{2}} \ = \ Z^{\text{FF}}_{B;S=0}(q,z)\ ,
 \eeqa
in agreement with our general prediction \eqref{main2}.

Although we have not been able to find a conclusive proof of
\eqref{main2} for $S\geq 2$, we wish to give some additional
supporting evidence. To this end, we need a few more details about
representations of \ospn\ and the corresponding values of the
quadratic Casimir element. The representations we are interested
in are labeled by integral dominant highest weights $\Lambda$ of
the form \beqa \Lambda&=&a_1\delta_1+a_2(\delta_1+\delta_2)+\cdots
+ a_{S}(\delta_1+\cdots \delta_S)+a_{S+1}\epsilon_1+\cdots
+a_{2S-1}(\epsilon_1+\cdots
\epsilon_{S-1})\nonumber\\[2mm]
&&+a_{2S}\frac{\epsilon_1+\cdots
+\epsilon_S-\epsilon_{S+1}}{2}+a_{2S+1}\frac{\epsilon_1+\cdots
+\epsilon_S+\epsilon_{S+1}}{2}\ ,  \eeqa where $\delta_i$ and
$\epsilon_j$ appear in the construction of the weight system of
\ospn\ and obey
$(\epsilon_i,\epsilon_j)=-(\delta_i,\delta_j)=\delta_{ij}$. The
numerical coefficients $a_i\in \mathbb{N}$ must moreover obey some
additional consistency conditions that can be found in
\cite{Frappat:1996pb}. The value of the quadratic Casimir in the
representation of weight $\Lambda$ can now be expressed in terms
of the coefficients $a_i$ as, \beqa
C_{\Lambda}&=&(\Lambda,\Lambda+2\rho)=-\sum_{i=1}^S\left(\sum_{j=i}^S
a_j-2i\right)\sum_{k=i}^S a_k+\frac{(a_{2S}-a_{2S+1})^2}{4}
\nonumber\\[2mm]
&&+\sum_{i=1}^S\left(\sum_{j=i}^{S-1}a_{S+j}+
\frac{a_{2S}+a_{2S+1}}{2}+2(S+1-i)\right)
\left(\sum_{k=i}^{S-1}a_{S+k}+\frac{a_{2S}+a_{2S+1}}{2}\right)
\nonumber \ . \eeqa The fundamental representation corresponds to
$a_1=1$ and $a_i=0$ for $i\neq 1$ so that $C_{\delta_1}=-(1-2) =1$
for all $S$. The value of the quadratic Casimir does not only
determine the deformation of conformal weights, see eq.\
\eqref{main2}. It is also needed to compute the conformal weight
\begin{equation}
\label{primary conformal dimension}
h_{\Lambda}\ =\ \frac{C_{\Lambda}}{2k}
\end{equation}
of fields that are primary with respect to the underlying affine
superalgebra at level $k$. In our case, the level $k$ must be set
to $k=1$, as before.

After this preparation we can begin to test equation \eqref{main2}.
Let us first try to recover the ground states of the free field
theory at $R=1$. It is clear that the vacuum state at $R=1$ is
obtained by deforming the unique \ospn\ invariant field with
weight $h=0$ at $R=\infty$. So, we can turn to the ground states
in the second sector of eq.\ \eqref{tZ2n} right away.  From
\eqref{Rinfty} we infer that the boundary PCM contains a single
field multiplet that transforms in the fundamental representation
with $\Lambda=\delta_1$ and has conformal weight $h=0$. Under the
proposed deformation, the conformal weight of this multiplet is
lifted from $h=0$ to $h=1/2$, since $C_{\delta_1}=1$. The latter
value agrees precisely with the ground state energy of the
corresponding affine representation when $k=1$ as given by
\eqref{primary conformal dimension}.

We want to go a little further and recover states in the $R=1$
model whose weight is one above the ground states. Let us pick,
for example, a multiplet that transforms on the representation
$\Lambda = 3\delta_1$. In the large radius limit, this
representation arises for the first time among the states of
weight $h=3$. In fact, in eq.\ \eqref{The spectrumn} terms
containing $y_1^3$ are multiplied by $q^3$ or higher powers of
$q$. Since $C_{3\delta_1}=3$, the proposal  \eqref{main2} tells us
that the weight of this multiplet gets deformed to
$h=3-\frac{3}{2}=\frac{3}{2}$. Hence, it should appear among the
first descendants of the sector over the fundamental
representation. Indeed, the irreducible representation with
highest weight $3\delta_1$ is contained in the tensor product of
the fundamental representation with the adjoint representation.
Thus, $Z^{\text{FF}}_{B;S}$ contains this representation with
$h=\frac{3}{2}$ exactly as predicted by eq.\ \eqref{main2}.

\section{Conclusions, open questions and outlook}

This work contains two central results. To begin with, we have
been able to compute the exact boundary spectrum of a volume
filling brane on the 3-dimensional supersphere \Ss\ for all values
of the curvature radius $R$. With a little bit of extra work it
should be possible to extend our formulas to higher dimensional
superspheres and also to other spectra, including the spectrum of
the bulk fields (see comments below). The second result concerns
the duality between the supersphere PCM and the \ospn GN model.
More specifically, we were able identify the spectrum at the
special point $R=1$ with that of a free field theory, namely of
the model \eqref{FFact} with a particular choice of boundary
conditions. This is consistent with a recent conjecture in
\cite{Candu:2008yw} and it provides extremely strong additional
support for the duality.

The supersphere \Ss\ and its higher dimensional generalizations
have been advocated in the past \cite{Mann:2004jr,Mann:2005ab} as
good toy models for the world-sheet description of string theory
on $AdS_5 \times \Sp^5$. Obviously, the defining equations for
both $AdS_5$ and $\Sp^5$ are very similar to our basic constraint
\eqref{constraint}. What is more important, however, is that the
world-sheet models for $AdS_5 \times \Sp^5 =
\bigl[\text{PSU}(2,2|4)/\text{SO}(1,4)\times\text{SO}(5)\bigr]_0$
and the supersphere theory give rise to continuous families of 2D
conformal field theories with many common features. In both cases,
the non-abelian global symmetries remain unbroken. On the other
hand, they are not enhanced into affine symmetries, at least not
for generic points in the moduli space. Consequently, it seems
reasonable to speculate briefly about possible lessons the
supersphere models might teach us for the world-sheet descriptions
of string theory in $AdS_5 \times \Sp^5$.\footnote{Similar remarks
  apply obviously to $AdS_4\times\mathbb{C}\mathbb{P}^3$.}

To begin with, it is certainly possible to determine the exact
spectrum of the free sigma model on the supercoset
$\text{PSU}(2,2|4)/\text{SO}(1,4)\times\text{SO}(5)$ at
$R=\infty$, much as this was done here for the supersphere. The
deformation of the spectrum away from $R=\infty$ cannot be as
simple as in the supersphere case. In fact, we know for sure that
there are some operators whose anomalous dimensions do not possess
a quasi-abelian dependence of the radius $R$ (or the 't Hooft
coupling). It might be interesting, however, to study whether
there is some subset of operators whose dimensions are given by
eq.\ \eqref{dimshiftSs}. Since we have nothing to say about this
right now, let us just imagine that in some way we were able to
deform the entire spectrum. Then we could start to look for
special values of the radius $R$ at which the spectrum contains
half-integer or integer values only. We know for sure that such a
point exists, namely the radius $R_0$ for with the string model
becomes dual to the free $N=4$ supersymmetric Yang-Mills theory.
One might hope that such a point is described by a free
world-sheet theory, just as it is the case for the superspheres.
In this sense, the dual of the free Yang-Mills theory would be the
analogue of the free GN model. If one found such strong-weak
coupling duality within the world-sheet description of strings in
$AdS$, it would reduce the AdS/CFT correspondence to a remaining
weak-weak coupling duality. World-sheet descriptions of weakly 
coupled gauge theory have appeared in the literature, see e.g.\ 
\cite{Gopakumar:2003ns,Aharony:2007fs} or the recent work 
\cite{Berkovits:2008qc} for two developments that seem 
relevant for what we have just outlined.

Finding an explicit action for such a free world-sheet model and
its deformation might have two interesting applications. To begin
with, it could provide a better starting point for the
quantization of the string theory on $AdS_5 \times \Sp^5$. In
fact, let us point out that our \OSPn-GN model is much simpler
than the original supersphere PCM: While the perturbative
expansion of the latter contains terms of any order in the basic
fields, the former has no terms beyond fourth order. Furthermore,
the perturbative expansion for the conjectured weakly coupled dual
of the strongly coupled $AdS_5\times \Sp^5$ sigma model could be
compared order by order to the perturbative expansion in the gauge
theory, see again \cite{Berkovits:2008qc}. One might even hope to
prove the AdS/CFT duality using such an intermediate world-sheet
model. Of course all this remains mere speculation for now. In
particular, it is clear that our analysis of supersphere models
exploited compactness of the
target's bosonic base. More work is necessary to include
non-compact targets such as $AdS_5 \times \Sp^5$ or
$AdS_4\times\mathbb{C}\mathbb{P}^3$.

After all these comments on possible implications for the AdS/CFT
correspondence, we would like to close with a few remarks on the
bulk spectrum of the supersphere models. The analysis of boundary
deformations in \cite{Quella:2007sg} puts much stress on the fact
that computations where only possible for very particular boundary
spectra. In fact, open strings had to be localized at one point in
a background in order to avoid running into mixing problems. For
the superspheres, similar issues do not arise. While \cite{Quella:2007sg}
focused on a bulk deformation preserving global left and right
transformations simultaneously, the current-current perturbation
\eqref{int} considered here is of a very different type. Since
the deforming operator does not involve any tachyonic vertex
operators, there is no mixing problem, neither for boundary
theories, nor even for the bulk. On the other hand, the
perturbation breaks the global bulk symmetry down to a single
diagonal action of the symmetry algebra. Therefore, it should
be possible to deform bulk spectra, but it might be more
difficult to identify the relevant \ospn\ action as we deform
from $R=1$ to $R=\infty$. We will return to these issues in a
future publication.

\bigskip\bigskip
\noindent {\bf Acknowledgments:} We would like to thank Thomas
Creutzig, Guiliano Niccoli, Peter R{\o}nne, J\"org Teschner, Alexei
Tsvelik and in particular Constantin Candu and Hubert Saleur for
numerous stimulating discussion and many useful comments. T.Q.\ and
V.S.\ are also grateful for the kind hospitality at the Isaac Newton
Institute and the inspiring atmosphere during its Workshop ``Strong
Fields, Integrability and Strings''.
The research of T.Q.\ is funded by a Marie Curie Intra-European
Fellowship, contract number MEIF-CT-2007-041765. We furthermore
acknowledge partial support from the EU Research Training Network
{\it Superstring theory}, MRTN-CT-2004-512194 and from {\it
ForcesUniverse}, MRTN-CT-2004-005104.
\newpage

\appendix

\section{Some aspects of the representation theory of $\mathbf{\text{OSP}(4|2)}$}

Our first appendix contains a number of basic notations and
results concerning the Lie superalgebra \osp. These are used
frequently in the main text. The complex superalgebra $g:=\osp$
may be realized as the set of matrices \beqa
\osp=\left\{\left(\begin{array}{cc}A & B\\ J_{2}B^t &
D\end{array}\right): A^t=-A\text{ and }
D^tJ_{2}=-J_{2}D\right\}\nonumber \eeqa with
$J_2=\left(\begin{smallmatrix}0 & -1 \\ 1 &
0\end{smallmatrix}\right)$ and the standard definition of graded
commutators. We have the usual separation of the
superalgebra into a bosonic
$g_{\bar{0}}=\text{sp}(2)\oplus\text{so}(4)\cong\text{sl}(2)\oplus
\text{sl}(2)\oplus\text{sl}(2)$
and a fermionic $g_{\bar{1}}$ subspace. In addition, the
superalgebra has a $\mathbb{Z}$-grading that is compatible with
its $\mathbb{Z}_2$ structure, i.e. $g=g_{-2}\oplus g_{-1}\oplus
g_0\oplus g_1\oplus g_2$, where the relation $[g_i,g_j]=g_{i+j}$
holds, with $g_0 \cong \text{so}(4)\oplus\text{gl}(1)$,
$g_{\bar{0}}=g_{-2}\oplus g_0\oplus g_2$ and
$g_{\bar{1}}=g_{-1}\oplus g_{1}$.

An integral dominant highest weight $\Lambda=(j_1,j_2,j_3)$ of
$g_{\bar{0}}$ is also one for the full superalgebra $g$ if it obeys
the consistency conditions:
\beqa
\label{consistency}
j_1=0\Rightarrow j_2=j_3=0\quad,\qquad j_1=\frac{1}{2}\Rightarrow j_2=j_3
\eeqa
where the first spin is related to the symplectic subalgebra and the
two others to the orthogonal one. The finite dimensional irreducible
representations $[\Lambda]$ of $g$ are constructed as follows. Taking
an irreducible highest weight representation $(\Lambda)$ of $g_0\cong
\text{so}(4)\oplus\text{gl}(1) $ with highest weight
$\Lambda=(j_1,j_2,j_3)$ associated to the highest weight vector
$v_{\Lambda}$, we set
\begin{align*}
M_{\Lambda}&\ =\ U(g)(E_1^-)^{2j_1+1}v_{\Lambda}\ \ ,&
K_{\Lambda}&\ =\ \left(\text{Ind}_p^g (\Lambda)\right)/M_{\Lambda}
\end{align*}
where $U(g)$ is the universal enveloping algebra of $g$, $E_1^-$
is the lowering operator of the symplectic subalgebra and
$p=g_0\oplus g_1\oplus g_2$. In the above equation, we have
considered the $g_{0}$-module $(\Lambda)$ as a $p$-module by
letting $g_{i}, i=1,2$ act trivially on it. The finite dimensional
representation $K_{\Lambda}$ is called the Kac module of $\Lambda$
and is generically irreducible. The set of Kac modules is divided
into typical and atypical ones. If the Kac module $K_{\Lambda}$ is
typical, then it is guaranteed to be irreducible. In this case we
define the simple module $[\Lambda]$ to be $K_{\Lambda}$. If,
however, one or more of the following atypicality conditions
\begin{equation}
  \begin{split}
\label{atypicality conditions}
2j_1&\ =\ -j_2-j_3\ ,\\
2j_1&\ =\ j_2+j_3+2\ ,\\
2j_1&\ =\ \pm(j_2-j_3)+1
  \end{split}
\end{equation}
hold, then $K_{\Lambda}$ is atypical and will generically contain a
maximal invariant subspace $I_{\Lambda}$ without being fully
reducible, i.e. it will contain indecomposable constituents. In those
cases, we set $[\Lambda]=K_{\Lambda}/I_{\Lambda}$. It can occur
however that $I_{\Lambda}=0$ even though $K_{\Lambda}$ is atypical.

The eigenvalue of the quadratic Casimir in the simple module
$[\Lambda]$ is given by the formula
\begin{equation}
\label{Casimir primary} C_2(\Lambda)\ = \
-4j_1(j_1-1)+2j_2(j_2+1)+2j_3(j_3+1)\ \ .
\end{equation}
In particular, $C_2(\Lambda)$ is always a square, i.e.\
$C_2(\Lambda) = k^2, k \in \mathbb{N}$, on atypical
representations $[\Lambda]$. The atypical weights
$\Lambda=(j_1,j_2,j_3)$ can be divided into blocks $\Gamma_k$,
such that weights in $\Gamma_k$ possess the same eigenvalue
$C_2(\Lambda)=k^2$ of the quadratic Casimir element. The
corresponding atypical labels can be listed explicitly
\cite{Germoni2000:MR1840448},
\begin{equation}
  \begin{split}
\Gamma_0&\ =\ \left\{\lambda_{0,0}=(0,0,0)\, ,\,
\lambda_{0,l}=\frac{1}{2}(l+1,l-1,l-1)\, ,\,  l\geq 1\right\}\\
\Gamma_k&\ =\ \left\{\lambda_{k,l}\, ,\,  l\in \mathbb{Z}\right\}
  \end{split}
\end{equation}
where
\beqa
\lambda_{k,l}=\left\{\begin{array}{lr}\frac{1}{2}(-l+2, -l-k, -l+k)&
    \text{ if }l\leq -k\\\frac{1}{2}(-l+1, l+k-1, -l+k-1) &\text{ if
    }-k+1\leq l\leq 0\\\frac{1}{2}(l+1, l+k-1, -l+k-1)& \text{ if
    }0\leq l\leq k-1\\ \frac{1}{2}(l+2,l+k,l-k)&  \text{ if } k\leq l
   \end{array}\right. \ \ .
\eeqa One sees easily, that the weights $\lambda_{k,-l}$ for
$k\geq 1$ may be obtained from $\lambda_{k,l}$ by simply
exchanging the second and the third Dynkin label. Furthermore, it
is possible to distinguish the weights $\lambda_{k,l}$ according
to the atypicality condition \eqref{atypicality conditions} they
obey. The only weight to fulfill the first condition is
$\lambda_{0,0}$. The weights belonging to the second condition are
$\lambda_{0,l}$ for $l\geq 1$ and $\lambda_{k,\pm l}$ for  $l\geq
k$. Finally, those the satisfy the last atypicality relation are
the $\lambda_{k,\pm l}$ for  $l< k$.

The only atypical Kac modules $K(\lambda_{k,l})$ which are
irreducible correspond to the weights $\lambda_{k,0}$ for $k\geq
0$ and to $\lambda_{0,1}$.  The indecomposable structure of the
remaining ones can be deciphered from the following diagram,
\begin{equation}
  \begin{split}
\label{Kac modules}
&K_{\lambda_{0,2}}:\ [\lambda_{0,2}]\longrightarrow [\lambda_{0,0}]\oplus [\lambda_{0,1}]\\
&K_{\lambda_{0,l}}:\ [\lambda_{0,l}]\longrightarrow [\lambda_{0,l-1}]\text{ for } l\geq 3 \\
&K_{\lambda_{k,l}}:\ [\lambda_{k,l}]\longrightarrow [\lambda_{k,l-1}]\text{ for } l\geq 1\\
&K_{\lambda_{k,l}}:\ [\lambda_{k,l}]\longrightarrow
[\lambda_{k,l+1}]\text{ for } l\leq -1\ \ .
  \end{split}
\end{equation}
The dimension of the typical Kac modules is \beqa
\dim[K_{(j_1,j_2,j_3)}] =16(2j_1-1)(2j_2+1)(2j_3+1) \eeqa whereas
the dimension of the atypical ones may be inferred from their
structure, together with the following formulas for the dimension
of the irreducible representations,
\begin{equation}
  \begin{split}
\label{atypical dimensions}
\dim [\lambda_{0,0}]&\ =\ 1\ ,\qquad \dim [\lambda_{0,1}]\ =\ 17\
,\qquad\dim [\lambda_{k,0}]\ =\ 4k^2+2\\
\dim [\lambda_{0,l}]&\ =\ (2l+1)\left[(2l+1)^2-3\right]\text{ for } l\geq 2\\
\dim [\lambda_{k,l}]&\ =\ (2l+1)\left[4(k^2-1)-(2l+1)^2+7\right]\text{ for } l\leq k-1\\
\dim [\lambda_{k,l}]&\ =\ (2l+3)\left[(2l+3)^2-4(k^2-1)-7\right]\text{ for } l\geq k
\end{split}
\end{equation}
where, of course, $\dim [\lambda_{k,-l}]=\dim [\lambda_{k,l}]$.
The decomposition of $K_{\Lambda}$ for $j_1\geq 1$, whether
typical or not,  into irreducible modules of the bosonic
subalgebra has been computed in \cite{VanDerJeugt:1985hq}. It
takes the form
\begin{equation}
  \begin{split}
\label{typical decomposition}
\left[K_{\Lambda}\right]_{g_{\bar{0}}}
  &\ \cong\ (j_1,j_2,j_3)\bigoplus_{\alpha,\beta=\pm\frac{1}{2}}
  (j_1-\frac{1}{2},j_2+\alpha,j_3+\beta)\\
  &\qquad\bigoplus_{\alpha=\pm 1}\big[(j_1-1,j_2+\alpha,j_3)\oplus(j_1-1,j_2,j_3+\alpha)\big]\oplus 2 (j_1-1,j_2,j_3)\\
&\qquad\oplus\bigoplus_{\alpha,\beta=\pm \frac{1}{2}}
(j_1-\frac{3}{2},j_2+\alpha,j_3+\beta)\oplus (j_1-2,j_2,j_3)\ \ .
\end{split}
\end{equation}
There are a few special cases for which the decomposition is not
generic. If $j_1\leq 2, j_2\leq 1$ or $j_3\leq 1$ then the above
decomposition formula must be truncated at the point where one ore
more of the labels become negative. Moreover, there are two cases
for which the multiplicity of the $(j_1-1,j_2,j_3)$ submodule has
to be changed. If $j_1=1, j_2>0, j_3>0$ or $j_1>1,j_2=0,j_3>0$ or
$j_1>1,j_2>0,j_3=0$, then this block will appear only once and if
both $j_2$ and $j_3$ are null, then it will not be present at all.

When $j_1=\frac{1}{2}$, the Kac modules $K_\Lambda$ with weight
$\Lambda$ obeying the consistency conditions \eqref{consistency}
are equal to the irreducible modules $\left[\frac{1}{2},
\frac{k}{2}, \frac{k}{2}\right]$ and they possess the following
structure \beqa \left[\frac{1}{2}, \frac{k}{2},
\frac{k}{2}\right]_{|g_{\bar{0}}}\cong \left(\frac{1}{2},
\frac{k}{2}, \frac{k}{2}\right)\oplus \left(0,
  \frac{k+1}{2}, \frac{k+1}{2}\right)\oplus \left(0, \frac{k-1}{2},
  \frac{k-1}{2}\right)\ \ .
\eeqa
Finally, the Kac module $K_{[0,0,0]}$ is trivial.

\section{Some useful identities}

In this appendix we collect a few definitions and identities that
we have employed to obtain the Casimir decompositions in sections
2.3 and 3.3. We also provide the first few terms in the Casimir
decomposition of the partition function $Z^{{\text{FF}}}_{B}$ for
$S=1$.

\subsection{Identities used in the Casimir decomposition}
To begin with, let us briefly recall the definition of Jacobi's
$\theta$ functions. In our conventions they are given by
\begin{equation} \label{thetaidentity}
\begin{split}
\theta_1(q|z)&=\, -i\sum_{r\in \mathbb{Z}+\frac{1}{2}}
(-1)^{r-\frac{1}{2}} z^r q^{\frac{r^2}{2}}\, =\, -i\,
z^{\frac{1}{2}} q^{\frac{1}{8}} \prod_{n=1}^{\infty} (1-q^n)(1-z
q^n)(1-z^{-1}
q^{n-1})\\[2mm]
\theta_2(q|z)&=\, \sum_{r\in \mathbb{Z}+\frac{1}{2}}z^r
q^{\frac{r^2}{2}}\, = \, z^{\frac{1}{2}}
q^{\frac{1}{8}}\prod_{n=1}^{\infty} (1-q^n)(1+z q^n)(1+z^{-1}
q^{n-1})\\[2mm]
\theta_3(q|z)&=\, \sum_{r\in \mathbb{Z}} z^r q^{\frac{r^2}{2}}\,
=\, \prod_{n=1}^{\infty} (1-q^n)\prod_{r\in
\mathbb{N}+\frac{1}{2}} (1+z q^r)(1+z^{-1}
q^{r})\\[2mm]
\theta_4(q|z)&=\, \sum_{r\in \mathbb{Z}}(-1)^r z^r
q^{\frac{r^2}{2}}\, =\, \prod_{n=1}^{\infty} (1-q^n)\prod_{r\in
\mathbb{N}+\frac{1}{2}} (1-z q^r)(1-z^{-1} q^{r})\ .
\end{split}
\end{equation}
The following two lemmata contain auxiliary formulas that are
needed to rewrite the partition function \eqref{The spectrum} in
terms of characters of \osp.
\begin{lemma}
\label{laurentidentity}
\begin{equation*}
\prod_{n=1}^{\infty} \frac{1}{(1-z q^n)(1-z^{-1} q^n)}\, =\,
\sum_{n\in \mathbb{Z}} z^n \sum_{m=0}^{\infty}(-1)^m
\frac{q^{\frac{m}{2}(m+2n+1)}-q^{\frac{m}{2}(m+2n-1)}}{\phi(q)^2}\ \ .
\end{equation*}
\end{lemma}
\begin{proof}
We assume that $|q|<|z|<1$, which is the relevant condition for the
above expansion to make sense. We want to find the coefficients
$f_l^N(q)$ in the relation
\begin{equation*}
\sum_{l\in \mathbb{Z}} f_l^N(q) z^l\ =\
\frac{1}{(1-z)\prod_{n=1}^N(1-z q^n) (1-z^{-1} q^n)} \ .
\end{equation*}
To do this, we multiply both sides by $z^{-k-1}$ and integrate
them over $z$ along a contour that surrounds zero in a
counterclockwise direction. In order to stay within the region
$|z|<1$ it must cling to the unit circle on the inside. The left
hand side of the previous equation gives us the coefficient
$f_k^N(q)$. The right hand side is zero for $z=0$ and the first
order poles that are encircled by the contour are at $z=q^n$ for
$n=1, \ldots , N$. Their residues are given by
\begin{equation*}
\lim_{z\rightarrow q^n} \frac{z^{-k-1}(z-q^n)}{(1-z) \prod_{l=1}^N
(1-z q^l) ( 1-z^{-1} q^l)}\ =\ \frac{(-1)^{n-1}
q^{\frac{n}{2}(n-2k-1)}}{\prod_{l=1}^{N+n}(1-q^l)
\prod_{l=1}^{N-n}(1-q^l)} \ .
\end{equation*}
If we finally remove our cutoff $N$ by sending $N\rightarrow
\infty$ we arrive at \begin{equation*}
\frac{1}{(1-z)\prod_{n=1}^{\infty}(1-z q^n) (1-z^{-1}
q^n)}=\sum_{k\in \mathbb{Z}} z^k \sum_{n=0}^{\infty}
\frac{(-1)^{n-1} q^{\frac{n}{2}(n-1-2k)}}{\phi(q)^2}\ \ .
\end{equation*}
Multiplying both sides by $1-z$ and using the lemma \ref{trivial
lemma} below to shuffle some minus signs around completes the
proof.
\end{proof}

\begin{lemma}
\label{trivial lemma} \beqa
&&\sum_{m=1}^{2n}(-1)^mq^{\frac{m(m-1)}{2}-mn}\ =\ 0\quad \quad
\mbox{\rm for } \ \ \ \
n\geq 1\nonumber\\[2mm]
&&\sum_{m=1}^{\infty}\sum_{s=-r}^r(-1)^mq^{\frac{m(m-1)}{2}-m(n+s)}
(1-q^m)\ =\
\sum_{m=1}^{\infty}\sum_{s=-r}^r(-1)^mq^{\frac{m(m-1)}{2}
-m(-n+s)}(1-q^m)\ .\nonumber \eeqa
\end{lemma}
\begin{proof}
The first equation is shown to be true by splitting the sum in
$\sum_{m=1}^{n}$ and $\sum_{m=n+1}^{2n}$ and showing that they are
equal up to a sign. The second equation then follows easily from
the first.
\end{proof}

There are a number of very simple auxiliary formulas that are
needed for the Casimir decomposition in section 2.3. Let us only
list two of them here
\begin{equation} \label{trick1}
\sum_{r=0}^{\infty}(-1)^r q^{\frac{r(r+2)}{4}}(1-q^{r+2})a_r\ =\
\sum_{r=0}^{\infty}(-1)^rq^{\frac{r(r+2)}{4}}(a_r-a_{r-2})
\end{equation}
\begin{equation}  \label{trick2}
\begin{split}
&\left(q^{(j_2-\frac{r}{2})^2}-q^{(j_2+\frac{r}{2}+1)^2}\right)
\left(q^{(j_3-\frac{r}{2})^2}-q^{(j_3+\frac{r}{2}+1)^2}\right)\ =\
q^{j_2(j_2+1)+j_3(j_3+1)}q^{\frac{r^2}{2}+r+1} \\[2mm]
&\hspace*{1cm} \times\
\left(q^{-(r+1)(j_2+j_3+1)}+q^{(r+1)(j_2+j_3+1)}-
q^{(r+1)(j_2-j_3)}-q^{-(r+1)(j_2-j_3)}\right)\ .
\end{split}
\end{equation}

\subsection{Casimir decomposition of $Z^{{\text{FF}}}_B$}

In section 3.3 we obtained closed formulas \eqref{tZdecform} and
\eqref{main} for the Casimir decomposition of the partition
function $Z^{\text{FF}}_B$. Since our expression for the
branching functions is a bit complicated, let us reproduce the
first few terms of the partition function explicitly, \beqa
Z^{\text{FF}}_{B; S=1}(q)\hspace*{-11mm}&&\hspace*{5mm}
=q^0\chi_{[0,0,0]}+q^{\frac{1}{2}}\chi_{[\frac{1}{2},0,0]}+
q^1\chi_{[1,0,0]}+q^{\frac{3}{2}}\left(\chi_{[\frac{3}{2},0,0]}+
\chi_{[\frac{1}{2},0,0]}\right)\nonumber\\&&+q^2\left(\chi_{[2,0,0]}
+\chi_{[1,0,0]}+\chi_{[\frac{1}{2},\frac{1}{2},\frac{1}{2}]}+
\chi_{[0,0,0]}\right)\nonumber \\ &&
+q^{\frac{5}{2}}\left(\chi_{[\frac{5}{2},0,0]}+
\chi_{[\frac{3}{2},0,0]}+\chi_{[1,\frac{1}{2},\frac{1}{2}]}+
2\chi_{[\frac{1}{2},0,0]}\right)\nonumber\\
&&+q^3\left(\chi_{[3,0,0]}+\chi_{[2,0,0]}+\chi_{[\frac{3}{2},
\frac{1}{2},\frac{1}{2}]}+4\chi_{[1,0,0]}+\chi_{[\frac{1}{2},
\frac{1}{2},\frac{1}{2}]}+\chi_{[0,0,0]}\right)\nonumber\\
&&+q^{\frac{7}{2}}\left(\chi_{[\frac{7}{2},0,0]}+
\chi_{[\frac{5}{2},0,0]}+\chi_{[2,\frac{1}{2},\frac{1}{2}]}+
3\chi_{[\frac{3}{2},0,0]}+2\chi_{[1,\frac{1}{2},\frac{1}{2}]}+
3\chi_{[\frac{1}{2},0,0]}\right)\nonumber\\
&&+q^4\left(\chi_{[4,0,0]}+\chi_{[3,0,0]}+\chi_{[\frac{5}{2},
\frac{1}{2},\frac{1}{2}]}+3\chi_{[2,0,0]}+2\chi_{[\frac{3}{2},
\frac{1}{2},\frac{1}{2}]}+\chi_{[1,1,0]}+\chi_{[1,0,1]}\right.
\nonumber\\&&\left.+6\chi_{[1,0,0]}+4\chi_{[\frac{1}{2},
\frac{1}{2},\frac{1}{2}]}+3\chi_{[0,0,0]}\right)\nonumber\\
&&+q^{\frac{9}{2}}\left(\chi_{[\frac{9}{2},0,0]}+\chi_{[\frac{7}{2},0,0]}
+\chi_{[3,\frac{1}{2},\frac{1}{2}]}+3\chi_{[\frac{5}{2},0,0]}+
2\chi_{[2,\frac{1}{2},\frac{1}{2}]}+\chi_{[\frac{3}{2},1,0]}\right.
\nonumber\\&&\left.+\chi_{[\frac{3}{2},0,1]}+5\chi_{[\frac{3}{2},0,0]}+
4\chi_{[1,\frac{1}{2},\frac{1}{2}]}+\chi_{[\frac{1}{2},1,1]}+
7\chi_{[\frac{1}{2},0,0]}\right)\nonumber\\
&&+q^5\left(\chi_{[5,0,0]}+\chi_{[4,0,0]}+\chi_{[\frac{7}{2},
\frac{1}{2},\frac{1}{2}]}+3\chi_{[3,0,0]}+2\chi_{[\frac{5}{2},
\frac{1}{2},\frac{1}{2}]}+\chi_{[2,1,0]}+\chi_{[2,0,1]}\right.
\nonumber\\&&\left.+5\chi_{[2,0,0]}+5\chi_{[\frac{3}{2},
\frac{1}{2},\frac{1}{2}]}+\chi_{[1,1,1]}+\chi_{[1,1,0]}+
\chi_{[1,0,1]}+14\chi_{[1,0,0]}+5\chi_{[\frac{1}{2},
\frac{1}{2},\frac{1}{2}]}+3\chi_{[0,0,0]}\right)\nonumber\\
&&+q^{\frac{11}{2}}\left(\chi_{[\frac{11}{2},0,0]}+
\chi_{[\frac{9}{2},0,0]}+\chi_{[4,\frac{1}{2},\frac{1}{2}]}+
3\chi_{[\frac{7}{2},0,0]}+2\chi_{[3,\frac{1}{2},\frac{1}{2}]}+
\chi_{[\frac{5}{2},1,0]}\right.\nonumber\\&&\left.+
\chi_{[\frac{5}{2},0,1]}+5\chi_{[\frac{5}{2},0,0]}+
5\chi_{[2,\frac{1}{2},\frac{1}{2}]}+10\chi_{[\frac{3}{2},0,0]}+
2\chi_{[\frac{3}{2},1,0]}+2\chi_{[\frac{3}{2},0,1]}+
\chi_{[\frac{3}{2},1,1]}\right.\nonumber\\ &&\left. +
8\chi_{[1,\frac{1}{2},\frac{1}{2}]}+\chi_{[\frac{1}{2},1,1]}+
11\chi_{[\frac{1}{2},0,0]}\right)\nonumber\\
&&+q^{6}\left(\chi_{[6,0,0]}+\chi_{[5,0,0]}+
\chi_{[\frac{9}{2},\frac{1}{2},\frac{1}{2}]}+3\chi_{[4,0,0]}+
2\chi_{[\frac{7}{2},\frac{1}{2},\frac{1}{2}]}+\chi_{[3,1,0]}
\right.\nonumber\\&&\left.+\chi_{[3,0,1]}+5\chi_{[3,0,0]}+
5\chi_{[\frac{5}{2},\frac{1}{2},\frac{1}{2}]}+11\chi_{[2,0,0]}+
2\chi_{[2,1,0]}+2\chi_{[2,0,1]}+\chi_{[2,1,1]}\right.\nonumber\\
 &&\left. +11\chi_{[\frac{3}{2},\frac{1}{2},\frac{1}{2}]}+
 2\chi_{[1,1,1]}+4\chi_{[1,1,0]}+4\chi_{[1,0,1]}+22\chi_{[1,0,0]}+
 13\chi_{[\frac{1}{2},\frac{1}{2},\frac{1}{2}]}+9\chi_{[0,0,0]}
 \right)\nonumber\\
&&+q^{\frac{13}{2}}\left(\chi_{[\frac{13}{2},0,0]}+
\chi_{[\frac{11}{2},0,0]}+\chi_{[5,\frac{1}{2},\frac{1}{2}]}+
3\chi_{[\frac{9}{2},0,0]}+2\chi_{[4,\frac{1}{2},\frac{1}{2}]}+
\chi_{[\frac{7}{2},1,0]}\right.\nonumber\\&&\left.+
\chi_{[\frac{7}{2},0,1]}+5\chi_{[\frac{7}{2},0,0]}+
5\chi_{[3,\frac{1}{2},\frac{1}{2}]}+11\chi_{[\frac{5}{2},0,0]}+
2\chi_{[\frac{5}{2},1,0]}+2\chi_{[\frac{5}{2},0,1]}+
\chi_{[\frac{5}{2},1,1]}\right.\nonumber\\ &&\left. +
11\chi_{[2,\frac{1}{2},\frac{1}{2}]}+2\chi_{[\frac{3}{2},1,1]}+
5\chi_{[\frac{3}{2},1,0]}+5\chi_{[\frac{3}{2},0,1]}+
16\chi_{[\frac{3}{2},0,0]}+15\chi_{[1,\frac{1}{2},\frac{1}{2}]}+
\chi_{[1,\frac{3}{2},\frac{1}{2}]}\right.\nonumber\\&&\left.+
\chi_{[1,\frac{1}{2},\frac{3}{2}]}+4\chi_{[\frac{1}{2},1,1]}+
21\chi_{[\frac{1}{2},0,0]}\right)+\ldots \nonumber\ \ . \eeqa One may
deform this expression to values $R \neq 1$ by means of the
formula \eqref{PFZR} at the end of section 3.4.

\section{\label{ap:Kac}Recombination of the bosonic characters}

Let $Z$ be a partition function with \osp\ symmetry. If we denote
the characters of the bosonic subalgebra by
$\chi^B_{(j_1,j_2,j_3)}(z_i)=\chi_{j_1}(z_1)\chi_{j_2}(z_2)
\chi_{j_3}(z_3)$, we can write the partition function as \beqa Z\
=\ \sum_{\lambda\in \cJ}
\chi^B_{\lambda}(z_1,z_2,z_3)\psi^B_{\lambda}(q)\ =\
\sum_{\lambda\in \cJ^{\prime}}
\chi^K_{\lambda}(z_1,z_2,z_3)\psi^K_{\lambda}(q) \eeqa where
$\cJ^{\prime}\subset \cJ$ is the set of labels in $\cJ = \{
(j_1,j_2,j_3); j_i = 0,1/2,1,3/2, \dots \}$ that are compatible
with the consistency conditions \eqref{consistency}. Here, the
first decomposition is in terms of bosonic characters while the
second one is based on the characters of Kac modules. In order to
find the relations between these two decompositions, we recall
that the roots of the four fermionic lowering operators in
$g_{-1}:=\text{osp}(4|2)_{-1}$ are \beqa
\alpha_1=\left(-\frac{1}{2}, \frac{1}{2}, \frac{1}{2}\right)\
\alpha_2=\left(-\frac{1}{2}, \frac{1}{2}, -\frac{1}{2}\right)\
\alpha_3=\left(-\frac{1}{2}, -\frac{1}{2}, \frac{1}{2}\right)\
\alpha_4=\left(-\frac{1}{2}, -\frac{1}{2}, -\frac{1}{2}\right)\ .
\eeqa Let us first discuss the generic label $\lambda =
(j_1,j_2,j_3)$ where either $j_1\geq \frac{3}{2}$, or $j_1=1$ and
$(j_2,j_3)\neq (0,0)$. In such cases we can write the
decomposition of the Kac module character $\chi^K_{\lambda}$ as
\beqa \chi^K_{\lambda}=\sum_{i=0}^4\sum_{\beta\in
\Lambda^{i}(g_{-1})} \chi^B_{\lambda+\beta} \eeqa where $\beta$ is
any of the weights that appear in the $i^{\text{th}}$ exterior
product $\Lambda^{i}(g_{-1})$ of $g_{-1}$. We also allow for
negative spins using the formal prescription
$\chi_{j}=-\chi_{-j-1}$. To treat the remaining cases with
$j_1\leq \frac{1}{2}$ we employ the formulas developed in
appendix A. Inserting the decomposition of Kac modules into the
partition function $Z$ leads to a formula that expresses the
bosonic branching functions $\psi^B_{\lambda}$ as sums of the
branching functions $\psi^K_{\mu}$. Our main aim is to invert this
relation, i.e.\ to determine the branching functions $\psi^K$ in
terms of $\psi^B$. To this end let us state a few basic properties
of $\psi^K$ that will be checked afterwards, once we have an
explicit formula,
  \beqa \label{symmetry assumption}
\psi^K_{[j_1,j_2,j_3]}\ =\ -\psi^K_{[j_1,-j_2-1,j_3]}\ =\
-\psi^K_{[j_1,j_2,-j_3-1]}
  \ \ .
\eeqa If we take this behavior of $\psi^K$ for granted the
decomposition formulas for the partition function $Z$ and of
$\chi^K$ in terms of bosonic characters imply, \beqa \label{bosonic
psi} \psi^B_{\lambda}=\sum_{i=0}^4\sum_{\beta\in
\Lambda^{i}(g_{-1})} \psi^K_{\lambda-\beta} \eeqa for all
$\lambda\in \cJ^{\prime}$. Inverting this expression leads to the
following result \beqa \label{supersymmetric psi}
\psi^K_{\Lambda}\ =\ \sum_{n=0}^{\infty}(-1)^n \sum_{\beta\in
\text{Sym}^{n}(g_{-1})}\psi^B_{\Lambda-\beta}\ \ . \eeqa To
establish formula \eqref{supersymmetric psi} we plug \eqref{bosonic
psi} into \eqref{supersymmetric psi}. Thereby we obtain \beqa
\label{auxiliary}
\psi^K_{\Lambda}\ =\
\sum_{i=0}^{\infty}(-1)^i\underbrace{\sum_{j=0}^4(-1)^j\sum_{\beta\in
\text{Sym}^{i-j}(g_{-1})}\sum_{\gamma\in
\Lambda^j(g_{-1})}\psi^K_{\Lambda-\beta-\gamma}}_{=0 \text{ if } i\neq 0}=\psi^K_{\Lambda} \ \ , \eeqa
thus showing that \eqref{supersymmetric psi} inverts \eqref{bosonic psi}.
In \eqref{auxiliary} we have set $\text{Sym}^n (V)=\emptyset$ if
$n<0$ and used the identity:
\begin{equation}
\label{auxiliary2}
\sum_{j=0}^4(-1)^j\sum_{\beta\in
\text{Sym}^{i-j}(V)}\sum_{\gamma\in
\Lambda^j(V)}c(\beta+\gamma)=0\ \ ,
\end{equation}
which is true for every four dimensional vector space $V$ and
every function $c$ as long as $i\geq 1$. To show
\eqref{auxiliary2}, we introduce the symbol $\ominus$ which is to
be understood as a sort of a negative of a direct sum as for
example in $A\oplus B \ominus B= A$. Then \eqref{auxiliary2} is
equivalent to $\bigoplus_{j=0}^4\ominus^j
\text{Sym}^{i-j}(V)\otimes\Lambda^j(V)=0$ if $i\geq 1$, which can
be shown using standard Young tableaux techniques. Denote a
tableau consisting of one single row with $m$ boxes by $1^m$ and a
tableau with one single
column of $n$ boxes\footnote{ Since we work with a four-dimensional
space $V$, $4^1=0^1$ must denote the trivial one-dimensional
space.} by $n^1$ and compute that $1^m\otimes n^1=1^mn^1\oplus
1^{m-1}(n+1)^1$ if $m\geq 1, n\geq 1, n\leq 4$. Thus \beqa
&&\bigoplus_{j=0}^4\ominus^j \text{Sym}^{i-j}(V)\otimes
\Lambda^j(V)=\bigoplus_{j=0}^4\ominus^j 1^{i-j}\otimes
j^1\nonumber\\&&=1^i\oplus \bigoplus_{j=1}^3\ominus^j\big[
1^{i-j}j^1\oplus1^{i-(j+1)}(j+1)^i\big]\oplus 1^{i-4}\otimes 4^1=0
\eeqa if $i\geq 1$. Thereby we have established that our
assumption \eqref{symmetry assumption} implies the result
\eqref{supersymmetric psi}.

In order to complete our proof of equation \eqref{supersymmetric
psi} we still need to verify our assumption \eqref{symmetry
assumption}. Let us observe that the bosonic branching functions
$\psi^B$ possess the same symmetry property, because, since the bosonic
characters $\chi^B$ are simply products of sl(2) characters
$\chi_j = - \chi_{-j-1}$, the identity \eqref{symmetry assumption}
holds trivially for $\psi^B$ instead of $\psi^K$. We can use this
fact to show  \beqa
\psi^K_{\omega_m(\lambda)}&=&\sum_{i=0}^{\infty}(-1)^i\sum_{\beta\in
\text{Sym}^{i}(g_{-1})}\psi^B_{\omega_m(\lambda)-\beta}=\sum_{i=0}^{\infty}
(-1)^i\sum_{\beta\in
\text{Sym}^{i}(g_{-1})}\psi^B_{\omega_m(\lambda-
\tilde{\omega}_m(\beta))}\nonumber\\&=&-\sum_{i=0}^{\infty}(-1)^i
\sum_{\beta\in
\text{Sym}^{i}(g_{-1})}\psi^B_{\lambda-\tilde{\omega}_m(\beta)}
=-\sum_{i=0}^{\infty}(-1)^i\sum_{\beta\in
\text{Sym}^{i}(g_{-1})}\psi^B_{\lambda-\beta}\ .  \eeqa  The
labels $\omega_2(\lambda)$ and $\tilde \omega_2(\lambda)$ were
introduced as $\omega_2(\lambda) = (j_1,-j_2-1,j_3)$ and
$\tilde{\omega}_2 (\lambda) =  (j_1,-j_2,j_3)$ for all $\lambda =
(j_1,j_2,j_3)$. Similar conventions apply to $\omega_3$ and
$\tilde \omega_3$.

As we have noted before, the functions $\psi^K_{\Lambda}$ can have
Laurent expansions with negative coefficients. Such negative
coefficients only appear in the atypical sector and they can be
traced back to the fact that we expanded the partition function
$Z$ in terms of `unphysical' characters of Kac modules rather than
through those of irreducible representations. The relation between
Kac modules and irreducible representation has direct implications
on the corresponding branching functions. In fact, the branching
functions  $\psi_\lambda$ that are defined through a decomposition
into characters of irreducible representations are related to the
branching functions $\psi^K$ by
$\psi_{[j_1,j_2,j_3]}(q)=\sum_{\Lambda} \psi_{\Lambda}^K(q)$. On
the right hand side the summation extends over all those
Kac modules $K_\Lambda$ that contain the irreducible
representation $[j_1,j_2,j_3]$ in their decomposition series. All
relevant decomposition series were spelled out in eq.\ \eqref{Kac
modules}. This gives
\begin{equation}
  \begin{split}
  \label{Formula for the irreducible characters}
\psi_{\lambda_{0,0}}(q)&\ =\ \psi_{\lambda_{0,0}}^K(q)+
\psi_{\lambda_{0,2}}^K(q)\\[2mm]
\psi_{\lambda_{0,l}}(q)&\ =\ \psi_{\lambda_{0,l}}^K(q)+
\psi_{\lambda_{0,l+1}}^K(q)\quad \quad \forall \ l\ \geq\ 1\\[2mm]
\psi_{\lambda_{k,0}}(q)&\ =\ \psi^K_{\lambda_{k,0}}(q)+
\psi^K_{\lambda_{k,1}}(q)+\psi^K_{\lambda_{k,-1}}(q)\quad
\quad \forall\  k\ \geq\  1 \\[2mm]
\psi_{\lambda_{k,l}}(q)&\ =\ \psi^K_{\lambda_{k,l}}(q)+
\psi^K_{\lambda_{k,l+1}}(q)\quad \quad \forall \ k\ \geq\  1\ ,\
l\ \geq\  1\\[2mm]
\psi_{\lambda_{k,l}}(q)&\ =\ \psi^K_{\lambda_{k,l}}(q)+
\psi^K_{\lambda_{k,l-1}}(q)\quad \quad \forall \ k\ \geq\  1\ , \
l\ \leq\  -1\ \ .
  \end{split}
\end{equation}
Let us stress that the branching functions $\psi_\Lambda(q)$ for
irreducible representations of \osp\ are guaranteed to have
non-negative integral coefficients.

\section{A free field construction for $\mathbf{\widehat{\text{osp}}(M|2N)_{1}}$}
\def\sa{i}
\def\sb{j}
\def\sc{k}
\def\sd{l}
\def\si{a}
\def\sj{b}
\def\sk{c}
\def\sl{d}

This appendix contains a free field construction of the affine
osp(M$|$2N) algebra at level $k=1$ in terms of free
fermions and several bosonic ghost systems. Let us decompose all
supermatrices $X\in\text{osp}(M|2N)$ into blocks according to
\beqa \label{Definition of osp} X\ =\
\left(\begin{array}{c|cc}\mathcal{E} & \bar{\mathcal{T}} &
\mathcal{T} \\\hline -\mathcal{T}^t & \mathcal{F} & \mathcal{G}\\
\bar{\mathcal{T}}^t & \bar{\mathcal{G}} &
-\mathcal{F}^t\end{array}\right) \eeqa where $\mathcal{E}$ is
antisymmetric and $\mathcal{G}, \bar{\mathcal{G}}$ are symmetric.
A basis for the various blocks in the supermatrix $X$ is provided
by \beqa &&\mathcal{E}_{\sa\sb}\ =\ e_{\sa\sb}-e_{\sb\sa}\qquad
\quad 1\, \leq\, \sa\, < \, \sb\, \leq \,
M\nonumber\\[2mm]
&&\mathcal{F}_{\si\sj}\ =\ e_{\si\sj}\qquad \quad  1\, \leq\,
\si\, ,\, \sj\, \leq\,  N
\nonumber\\[2mm]
&& \mathcal{G}_{\si\sj}\ =\ \bar{\mathcal{G}}_{\si\sj}\ =\
e_{\si\sj}+e_{\sj\si}\qquad \quad
1\, \leq\,  \si\, \leq\,  \sj\, \leq\,  N\nonumber\\[2mm]
&&\mathcal{T}_{\sa\si}\ =\ \bar{\mathcal{T}}_{\sa\si}\ =\
e_{\sa\si}\qquad \quad  1\, \leq\,  \sa\, \leq\,  M\, ,\, 1\,
\leq\, \si \, \leq\,  N \eeqa where $e_{mn}$ are elementary
matrices. The matrices we have just introduced describe the
various blocks in the supermatrix $X$. We agree to denote by
$E_{\sa\sb}$ the supermatrix of the form \eqref{Definition of osp}
where ${\mathcal{E}}$ is given by $\mathcal{E}_{\sa\sb}$ and all
other blocks vanish. The basis elements
$F_{\si\sj},G_{\si\sj},\bar{G}_{\si\sj}, T_{\sa\si},
\bar{T}_{\sa\si}$ are defined similarly.

Now let us introduce $M$ free fermions $\psi_\sa$ and $2N$ bosons
$\beta_\si, \gamma _\si$ with the following basic operator
products, \beqa \psi_\sa(z)\psi_\sb(w)\, \sim\,
\frac{\delta_{\sa\sb}}{z-w}\ ,\qquad \beta_\si(z)\gamma_\sj(w)\,
\sim\, -\gamma_\si(z)\beta_\sj(w)\, \sim \,
\frac{\delta_{\si\sj}}{z-w}\ . \eeqa We can define the free field
representation of the osp(M$|$2N) current algebra through
\begin{equation}
  \begin{split}
E_{\sa\sb}(z)&\ =\ (\psi_\sa\psi_\sb)(z)\ ,\qquad
  F_{\si\sj}(z)\ =\ -(\beta_\si\gamma_\sj)(z)
\nonumber\\[2mm]
G_{\si\sj}(z)&\ =\ (\beta_\si\beta_\sj)(z)\ ,\qquad \
\bar{G}_{\si\sj}(z)\ =\ -(\gamma_\si\gamma_\sj)(z)
\nonumber\\[2mm]
T_{\sa\si}(z)&\ =\ i(\psi_\sa\beta_\si)(z)\ ,\qquad
\bar{T}_{\sa\si}(z)\ =\ -i(\psi_\sa\gamma_\si)(z)\ \ .
\end{split}
\end{equation}
The invariant bilinear form for osp(M$|$2N) is
$(X,Y)=\frac{1}{2}\text{str}(XY)$. On the basis elements it takes
the following from  \beqa
(E_{\sa\sb},E_{\sc\sd})&=&-\delta_{\sa\sc}\delta_{\sb\sd}\quad
\sa<\sb\text{ and } \sc<\sd
\nonumber\\[2mm]
(F_{\si\sj},F_{\sk\sl})&=&-\delta_{\si\sl}\delta_{\sj\sk}\nonumber\\[2mm]
(G_{\si\sj},\bar{G}_{\sk\sl})&=&-\delta_{\si\sk}\delta_{\sj
\sl}\quad \text{for } \si\neq \sj \text{ and } \sk\neq \sl \quad
(G_{\si\si}, \bar{G}_{\sj\sj})=
-2\delta_{\si\sj}\nonumber\\[2mm]
(T_{\sa\si},\bar{T}_{\sb\sj})&=&\delta_{\sa\sb}\delta_{\si\sj}\ \
. \eeqa With the help of this form and assuming that $M\neq 2N+1$,
the holomorphic part of the energy momentum tensor is given by the
Sugawara construction \beqa
T(z)&=&\frac{(J^{\mu}J_{\mu})(z)}{2(k+g^{\vee})}=\frac{1}{2(k+g^{\vee})}
\Big[-\sum_{\sa<\sb=1}^M(E_{\sa\sb}^2)-\sum_{\si,\sj=1}^N
(F_{\si\sj}F_{\sj\si})-\sum_{\si<\sj=1}^N\big(\left\{G_{\si\sj},\bar{G}_{\si\sj}\right\}\big)
\nonumber\\[2mm] &&-\frac{1}{2}\sum_{\si=1}^N\big(\left\{G_{\si\si},\bar{G}_{\si\si}
\right\}\big)-\sum_{\sa=1}^M\sum_{\si=1}^N\big(\left[T_{\sa\si},\bar{T}_{\sa\si}\right]
\big)\Big]\nonumber\\[2mm]
&=&-\frac{1}{2}\sum_{\sa=1}^M(\psi_\sa\partial
\psi_\sa)+\frac{1}{2}\sum_{\si=1}^N\big((\beta_\si\partial\gamma_\si)-
(\gamma_\si\partial\beta_\si)\big) \eeqa Here, the dual Coxeter
number is given by $g^{\vee}=M-2N-2$ and the value of the level is
$k=1$. The central charge of the system is easily seen to take the
value $c=\frac{M}{2}-N$.

Let us now introduce the involutive automorphism $\Omega$ such
that the fixed point set $\{X\in\text{osp}(M|2N)| \Omega(X)=X\}$ is
isomorphic to $\text{osp}(M-1|2N)$. On the basis we introduced above,
$\Omega$ acts non-trivially only on $E_{\sa\sb}, T_{\sa\si},
\bar{T}_{\sa\si}$. In fact, it multiplies all operators with
$\sa=1$ by $-1$ and leaves the others invariant. If we denote the
anti-holomorphic fields corresponding to $\psi_\sa, \beta_\si
,\gamma_\si$ by $\bar{\psi}_\sa,
\bar{\beta}_\si,\bar{\gamma}_\si$, the deformation operator
$J^{\mu}\Omega(\bar{J}_{\mu})$ can then be written as \beqa
\label{deformation operator}
J^{\mu}\Omega(\bar{J}_{\mu})&=&-\sum_{\sa<\sb=1}^M \varpi_\sa
(\psi_\sa \psi_\sb)(\bar{\psi}_\sa \bar{\psi}_\sb)-
\sum_{\si,\sj=1}^N(\beta_\si
\gamma_\sj)(\bar{\beta_\sj}\bar{\gamma}_\si)
\nonumber\\[2mm]&&+\!\!\sum_{\si<\sj=1}^N\left[(\beta_\si\beta_\sj)
(\bar{\gamma}_\si\bar{\gamma}_\sj)+(\gamma_\si\gamma_\sj)
(\bar{\beta}_\si\bar{\beta}_\sj)\right]+
\frac{1}{2}\sum_{\si=1}^N\left[(\beta_\si\beta_\si)
(\bar{\gamma}_\si\bar{\gamma}_\si)+(\gamma_\si\gamma_\si)
(\bar{\beta}_\si\bar{\beta}_\si)\right] \nonumber\\[2mm]
&&-\sum_{\sa=1}^M\sum_{\si=1}^N \varpi_\sa
\left[(\psi_\sa\beta_\si)(\bar{\psi}_\sa\bar{\gamma}_\si)-
(\psi_\sa\gamma_\si)(\bar{\psi}_\sa\bar{\beta}_\si)\right]
\nonumber\\[2mm]
&=&\frac{1}{2}\left[\sum_{\sa=1}^M\varpi_{\sa}\psi_\sa
\bar{\psi}_\sa+\sum_{\si=1}^N\big(\gamma_\si\bar{\beta}_\si
-\beta_\si\bar{\gamma}_\si\big)\right]^2 \eeqa where
$\varpi=(-1, 1,\ldots , 1)$. In order for the last line of
\eqref{deformation operator} to make sense, we need to first
expand the square and then bring all the fields in the standard normal
ordering.

\def\cprime{$'$} \def\cprime{$'$}
\providecommand{\href}[2]{#2}\begingroup\raggedright\endgroup

\end{document}